\documentclass[%
reprint,
superscriptaddress,
amsmath,amssymb,
aps,
prb,
floatfix,
]{revtex4-1}
\usepackage[version=3]{mhchem}
\usepackage{graphicx}
\usepackage{dcolumn}
\usepackage{bm}

\usepackage[utf8]{inputenc}
\usepackage[T1]{fontenc}
\usepackage{mathptmx}
\usepackage{bbold}
\usepackage{float}
\usepackage{booktabs}
\usepackage{multirow}
\usepackage{amsmath}
\usepackage{amssymb}
\usepackage{mathtools}
\usepackage{xcolor}

\begin{document}

\title[Halogen vacancy migration in CsPbBr$_3$]{Halogen Vacancy Migration at Surfaces of CsPbBr$_3$ Perovskites: Insights from Density Functional Theory}
\author{Raisa-Ioana Biega}
\affiliation{Institute of Physics, University of Bayreuth, Bayreuth 95440, Germany}
\author{Linn Leppert}
\email{l.leppert@utwente.nl}
\affiliation{MESA+ Institute for Nanotechnology, University of Twente, 7500 AE Enschede, The Netherlands}
\affiliation{Institute of Physics, University of Bayreuth, Bayreuth 95440, Germany}

\begin{abstract}
Migration of halogen vacancies is one of the primary sources of phase segregation and material degradation in lead-halide perovskites. Here we use first principles density functional theory to compare migration energy barriers and paths of bromine vacancies in the bulk and at a (001) surface of cubic CsPbBr$_3$. Our calculations indicate that surfaces might facilitate bromine vacancy migration in these perovskites, due to their soft structure that allows for bond lengths variations larger than in the bulk. We calculate the migration energy for axial-to-axial bromine vacancy migration at the surface to be only half of the value in the bulk. Furthermore, we study the effect of modifying the surface with four different alkali halide monolayers, finding an increase of the migration barrier to almost the bulk value for the NaCl-passivated system. Migration barriers are found to be correlated to the lattice mismatch between the CsPbBr$_3$ surface and the alkali halide monolayer. Our calculations suggest that surfaces might play a significant role in mediating vacancy migration in halide perovskites, a result with relevance for perovskite nanocrystals with large surface-to-volume ratios. Moreover, we propose viable ways for suppressing this undesirable process through passivation with alkali halide salts.
\end{abstract}
\maketitle
\section{Introduction}
Halide perovskites are exciting materials with exceptional optoelectronic properties, wide tunability, and a broad range of applications spanning solar cells \cite{Kojima2009, Liu2013, Stranks2015b}, light-emitting diodes (LEDs) \cite{Tan2014, Ling2016}, photo-detectors \cite{Zhang2018, Saidaminov2017, Ding2017} and X-ray scintillators \cite{Zhou2021}. Solar cells based on lead-halide perovskites ABX$_3$ with A=(CH$_3$NH$_3$)$^+$ (methylammonium, MA$^+$), (NH$_2$CHNH$_2$)$^+$ (formamidinium, FA$^+$), Cs$^+$, B=Pb$^{2+}$, and X=Cl$^-$, Br$^-$, I$^-$, can be processed at low-temperature, and have exceeded power conversion efficiencies of 25\% \cite{nrel2021}. Yet, commercialisation of perovskite-based solar cells and other devices is hampered by the lack of stability of the perovskite absorbers towards moisture, oxygen, light, heat, and electric fields \cite{Senocrate2019}. Various strategies have been applied to improve the stability of these materials, including encapsulation, (partial) replacement of the A site cation \cite{Saliba2016}, and passivation \cite{Schileo2020}. All-inorganic lead-halide perovskites CsPbX$_3$ have seen their own surge of interest, in particular because colloidal CsPbX$_3$ nanocrystals can exhibit very high photoluminescence quantum yields, with band gap energies and emission spectra tunable over the entire visible spectral region \cite{Huang2016a}. However, even all-inorganic halide perovskites can exhibit poor stability under electric fields \cite{He2018}. Material degradation and phase separation in both organic-inorganic and all-inorganic halide perovskites have been attributed to the migration of mobile ionic species \cite{Zhang2019}.

Ion migration in halide perovskites has been studied since the 1980s \cite{Mizusaki1983}. The dominant migrating species in these materials are halogen ions \cite{Mosconi2016b, Meloni2016, Luo2017, Senocrate2017}, mediated by the presence of halogen defects. The mechanism of halogen migration has been studied both experimentally \cite{Mizusaki1983, Narayan1987, Eames2015, Yuan2016, Yang2015, Lee2019} and using first principles simulation techniques such as density functional theory (DFT) \cite{Eames2015, Egger2015, Azpiroz2015, Haruyama2015, Meloni2016, Oranskaia2018}. And while reported activation energies for these migration processes span a wide range from $\sim$0.1 to $\sim$1.0\,eV \cite{Eames2015, Mosconi2016a, Mosconi2016b, Meloni2016, Luo2017, Senocrate2017, Chen2019a}, there is a consensus that halogen migration is the primary channel for the ionic conductivity observed in halide perovskites. The large spread of the experimental values has been linked to synthesis conditions, experimental techniques, and the role of grain sizes for defect formation in polycrystalline thin films \cite{Futscher2019, Zhang2020b}.

Prior first principles calculations of ion migration in both organic-inorganic and all-inorganic halide perovskites have, with a few exceptions, focused on ion migration in the bulk. However, in perovskite nanocrystals, ion migration at surfaces is expected to play an increasingly large role with decreasing particle size. Furthermore, in halide perovskite thin films, Kelvin probe force microscopy was used to demonstrate that ion migration is dominant at grain boundaries \cite{Xing2016, Yun2016}. More recently, the effect of surfaces and grain boundaries has also been explored via first principles calculations \cite{Oranskaia2018, Meggiolaro2019}. However, these computational studies have provided a mixed picture: Meggiolaro et al.~reported that migration energy barriers of interstitial iodine vacancies are little affected by surfaces \cite{Meggiolaro2019}. On the other hand, Oranskaia et al.~showed a clear effect of the surface on Br vacancy and interstitial migration in MAPbBr$_3$ and FAPbBr$_3$ \cite{Oranskaia2018}. For both materials, the activation energies of Br vacancy migration were computed to be 0.3\,eV lower than in the bulk. The variation in calculated results can have a number of sources such as differences in the applied level of theory, for example the choice of the DFT exchange-correlation functional and the method used for calculating migration barriers. A complication that is particular to the organic-inorganic perovskites, is that in the majority of DFT calculations, the rotational dynamics of the organic cation at room and higher temperatures are not taken into account. Instead structural models with fixed orientations of the molecular moieties are used, leading to significant differences in the potential energy landscape depending on the choice of molecular orientation \cite{Quarti2014}. Indeed, Oranskaia et al.~also showed that activation energies for Br migration significantly depend on the orientation of the organic moiety. The uncertainties associated with the choice of a suitable structural model for organic-inorganic perovskites at elevated temperatures, and the important role of surfaces in all-inorganic halide perovskite nanocrystals, are motivating our first principles study of bromine vacancy migration in the bulk and at a surface of cubic CsPbBr$_3$.

Our DFT calculations of vacancy-mediated bromine migration paths and energy barriers in CsPbBr$_3$, show a significant dependence on the presence of a surface. We find that the migration barrier in the bulk is about twice as large as the one at the surface. We show that variations of the Br migration barriers are correlated with variations in the Pb-Br bond length of bonds in the vicinity of the vacancy: halide migration at the surface is facilitated by the larger structural flexibility of the surface as compared to the bulk. Furthermore, migration paths considerably differ between the surface and the bulk, which can also be traced back to more flexible bonds at the surface. Finally, we study the effect of surface modification with alkali halide monolayers, demonstrating that a NaCl passivation layer leads to an increase of the migration energy of Br vacancy migration to almost the value in the bulk of the material.

\section{Methods}\label{methods}
\begin{figure}[ht]
    \centering
    \includegraphics[width=\columnwidth]{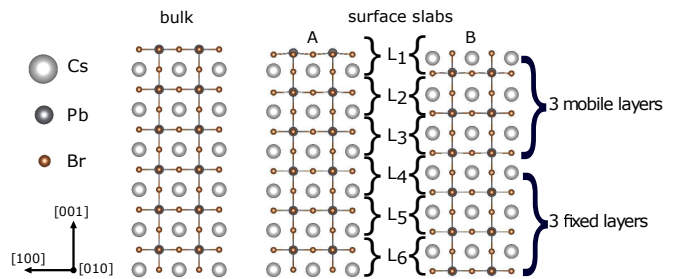}
    \caption{Bulk and slab supercells with A (PbBr$_2$) and B (CsBr) terminations. The label L$_i$ ($i=1-6$) enumerates layers in the slab structure. Surface slabs are separated by 30\,\AA~of vacuum along the [001] direction.}
    \label{fig1:unit-cell-setup}
\end{figure}

CsPbBr$_3$ is orthorhombic with $Pbnm$ symmetry at room temperature and undergoes two successive phase transitions to tetragonal ($P4/mbm$) at 88\textdegree C and to cubic ($Pm\bar{3}m$) at 130\textdegree C \cite{Hirotsu1974, Stoumpos2013}. Figure~\ref{fig1:unit-cell-setup} depicts the bulk and surface slab structures of cubic CsPbBr$_3$ used in this work. For constructing these structural models, we first performed a geometry optimization starting from the experimental high temperature crystal structure of CsPbBr$_3$ with $Pm\bar{3}m$ symmetry using DFT within the PBEsol approximation \cite{Perdew2008} as implemented in the Vienna Ab$-$initio Software Package (VASP) \cite{Kresse1993, Kresse1996}. The resulting optimized lattice parameter of 5.86\,\AA~is in very good agreement with the experimental value from X-ray diffraction \cite{Lopez2020}. To model the surface of CsPbBr$_3$, we designed two (001) surface slab supercells with distinct surface terminations by repeating the primitive $Pm\bar{3}m$ unit cell with PBEsol-optimized lattice parameters twice along the [100] direction, once along the [010] direction and six times along the out-of-plane [001] direction, with the bottom three layers fixed to bulk positions and the top three layers fully mobile. Unless otherwise specified, all our calculations were performed with this unit cell setup. Surface A is PbBr$_2$-terminated and surface B is CsBr-terminated. The surface energy is converged to within 25\,meV with respect to the slab thickness. To avoid spurious interactions between periodic images, we inserted 30\,\AA~of vacuum along the (001) direction. The bulk system features the same number of layers without vacuum. We used the projector augmented wave (PAW) method \cite{Kresse2014}, a cutoff energy for the plane-wave expansion of 300\,eV and a k-point grid with $4\times4\times4$ points for the bulk and $4\times4\times1$ points for the slab system. For geometry optimizations, we used a convergence criterion of 0.05\,eV/\AA. In all structural optimizations, the volume and shape of the unit cells was kept fixed.

We computed migration paths and migration energies using the climbing-image Nudged Elastic Band (cNEB) approach \cite{Henkelman2000}, an optimization method for identifying the minimum energy path between a given initial and final state. We used three cNEB images, i.e., intermediate structure snapshots between the initial and the final state of the system, to simulate the migration of the Br vacancy. In the cNEB method, the images along the reaction path are optimized such that the highest energy image is driven up to the saddle point. The migration barrier represents the amount of energy necessary for an ion or a defect to move from the initial to the final state and is calculated as the difference between the energy of the saddle point and the energy of the initial state of the transition.

\section{Results and discussion}
\subsection{Structural changes upon vacancy formation}
\begin{figure*}[ht]
    \centering
    \includegraphics[width=0.9\textwidth]{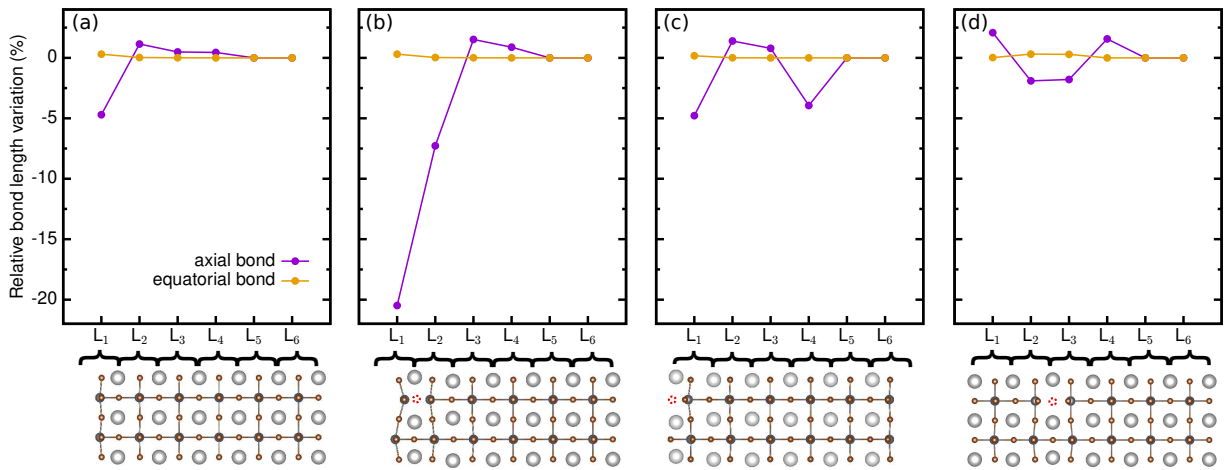}
    \caption{(a) Relative variation of the bond lengths in the surface slab with respect to the undistorted bulk bond lengths. Relative bond length variation when a Br vacancy is introduced (b) at surface A; (c) at surface B; (d) in the bulk. The geometry optimized structure is shown below each panel, with the position of the Br vacancy highlighted in red.}
    \label{fig2:bonds-variations}
\end{figure*}

We start by performing geometry optimizations of the bulk and A- and B-terminated surfaces shown in Figure~\ref{fig1:unit-cell-setup}. As expected, the bulk system remains unaffected by further geometry optimization, while the slab structure is compressed at the surface, with axial Pb-Br bonds by more than 4\,\% shorter than in the bulk and almost unchanged equatorial bonds. The average relative variation of Pb-Br bonds per layer as compared to the bulk Pb-Br bond length of 2.93\,\AA~is shown in Figure~\ref{fig2:bonds-variations}(a), where we have averaged over all axial and equatorial Pb-Br bonds (per layer), respectively.
A DFT study on intrinsic point defects in CsPbBr$_3$ showed that under Br-poor growth conditions, bromine vacancies have the lowest formation energy among all possible point defects \cite{Kang2017}. In our surface slab structures, a Br vacancy can occupy three symmetry-inequivalent positions in each layer of surface A and B: axial, equatorial along the [100] direction, and equatorial along the [010] direction. Note, that these three vacancy positions also have different energies in our structural model for the bulk, which is an artifact of our asymmetric unit cell. We define the formation energy of a vacancy in the slab ($E_f^{slab}$) and in the bulk ($E_f^{bulk}$) as the difference between the energies of the unit cell with and without vacancy. In Table~\ref{tbl1:formation_energy}, we report the binding energy $E_B = E_f^{bulk} - E_f^{slab}$ to quantify by how much a vacancy prefers to bind to the surface as compared to the bulk. $E_B$ is largest in L$_1$ and converges to zero in subsequent layers, in agreement with results for MAPbI$_3$ by Meggiolaro et al.~\cite{Meggiolaro2019}, suggesting that the surface is more prone to defects. We further find that binding to surface A is preferred over binding to surface B, in line with observations of iodine vacancy clustering at MAI-terminated surfaces in MAPbI$_3$ \cite{Zhang2019a}. For completeness, we also report $E_B$ for the two equatorial vacancy positions and note that this value is significantly larger for the vacancy along [010] because of our $2\times1\times6$ unit cell setup. In the following, we will only discuss ion migration between axial vacancies.

\begin{table}[ht]
    \centering
    \caption{Binding energies $E_B$ of Br vacancies to the surface as defined in the text.}
    \label{tbl1:formation_energy}
    \begin{tabular}{ccccc}
    \toprule
    Position & Layer & Termination & 
$E_B$ (eV)\\
    \midrule
     \multirow{4}{*}{axial} & \multirow{2}{*}{1} & A & 0.42 \\
                           &                    & B & 0.23 \\
                           & \multirow{2}{*}{2} & A & 0.22 \\
                           &                    & B & 0.23 \\
                           & \multirow{2}{*}{3} & A & 0.05 \\
                           &                    & B & 0.02 \\
                           \multirow{2}{*}{equatorial along [100]} & \multirow{2}{*}{1} & A & 0.35 \\
                           &                 & B & 0.15 \\
                           \multirow{2}{*}{equatorial along [010]} & \multirow{2}{*}{1} & A & 2.57 \\
                           &                 & B & 2.43 \\
    \bottomrule
    \end{tabular}
\end{table}

In Figure~\ref{fig2:bonds-variations}(b) and (c) we show the average Pb-Br bond length variation with respect to the undistorted bulk bond length upon introduction of the axial Br vacancy and geometry optimization. The creation of a Br vacancy at surface A leads to severe distortions of the system, featuring axial Pb-Br bonds reduced by up to 20\,\%. In contrast, introducing a vacancy at surface B leads to smaller variations and a less distorted structure. 
We have also calculated the variation of the Pb-Br bond length for a Br vacancy generated in the deeper lying surface layers, and find that even though the absolute value of the bond length variation differs for the two surface terminations, in both cases the variation in axial bond length is $\sim$5 times larger than the variation in equatorial bonds. Furthermore, the deeper the vacancy is created, the less compressed the structure is at the surface. In Figure~\ref{fig2:bonds-variations}(d) we show that formation of a Br vacancy within the bulk has similar consequences, i.e., a bond length compression in the vicinity of the vacancy. However, the distortions in the bulk are highly suppressed due to a more rigid structure, with fewer degrees of freedom in comparison with the surface slabs, leading to a zero average variation of the bond lengths when averaging over all mobile layers. Note that the large compression of more than 20\% in L$_1$ of surface A is an artifact of the asymmetric unit cell. In a $2\times2\times6$ cell, the compression of axial bonds is smaller than in the $2\times1\times6$ cell, with a relative bond length compression of 4.6\% in L$_1$ of surface A without a vacancy, 3.4\% in L$_1$ of surface A with a vacancy and hardly any variation with respect to the undistorted bulk for the case of a vacancy in the bulk. However, the trends for subsequent layers are similar to the $2\times1\times6$ unit cell.

\subsection{Br vacancy migration in CsPbBr$_3$}
Next, we use the cNEB method to determine the energy barrier of Br vacancy migration between two adjacent axial vacancy positions at both surfaces and for vacancy migration in layers L$_2$ and L$_3$ of the A-terminated surface. The migration barrier is calculated as the difference between the total energies of the initial state and the saddle point. Table~\ref{tbl2:activation_energy} summarizes our results. For the bulk structure, we compute a migration barrier of 0.65\,eV. Our specific unit cell setup is by design not suitable for direct comparison with experimental results or bulk calculations of halide migration in symmetric structural models, such as the one used in Ref.~\cite{Zhang2020}. However, our setup allows us to realize the same defect concentration and in-plane boundary conditions in the bulk and slab unit cells, and hence compare trends. It is worth mentioning that our calculations are in good agreement with the experimental values ranging between 0.72 and 0.66\,eV reported in the literature \cite{Mizusaki1983, Narayan1987}. However, our result is 140\,meV larger than the calculated migration barrier reported by Zhang et al.~\cite{Zhang2020}. This discrepancy may be explained based on different structures, defect concentrations, and approximations for the exchange correlation energy (Ref.~\cite{Zhang2020} uses the orthorhombic phase of CsPbBr$_3$ and the PBE approximation). Furthermore, we expect that our calculations represent an upper bound on the migration barriers, since we are neglecting the large, anharmonic vibrations reported for CsPbBr$_3$ and other halide perovskites at room and higher temperatures \cite{Yaffe2017}.

\begin{table*}[htb]
\caption{Calculated energies (in eV) of Br vacancy migration across two adjacent axial Br positions, and deviation (in \AA) from the straight migration path in CsPbBr$_3$ perovskite}
\label{tbl2:activation_energy}
\centering
\begin{tabular}{cccccc}
\toprule
System & Mobile layers & Termination & Layer & Migration energy (eV) & $\delta$ (\AA)\\
\midrule
 bulk                 & &                    &      & 0.65 & 0.13\\\midrule
\multirow{5}{*}{slab} &\multirow{5}{*}{3} & \multirow{3}{*}{A} & 1 & 0.40 & 1.24\\
                      &                   &                    & 2 & 0.29 & 0.94\\
                      &                   &                    & 3 & 0.30 & 1.04\\
                      &                   & \multirow{2}{*}{B} & 1 & 0.31 & 0.77\\
                      &                   &                    & 2 & 0.30 & 0.67\\\midrule
\multirow{4}{*}{slab} &\multirow{4}{*}{4} & \multirow{4}{*}{A} & 1 & 0.38 & 1.24\\
                      &                   &                    & 2 & 0.26 & 1.04\\
                      &                   &                    & 3 & 0.27 & 1.06\\
                      &                   &                    & 4 & 0.29 & 1.10\\
\bottomrule\end{tabular}
\end{table*}

\begin{figure}[ht]
    \centering
    \includegraphics[width=0.8\columnwidth]{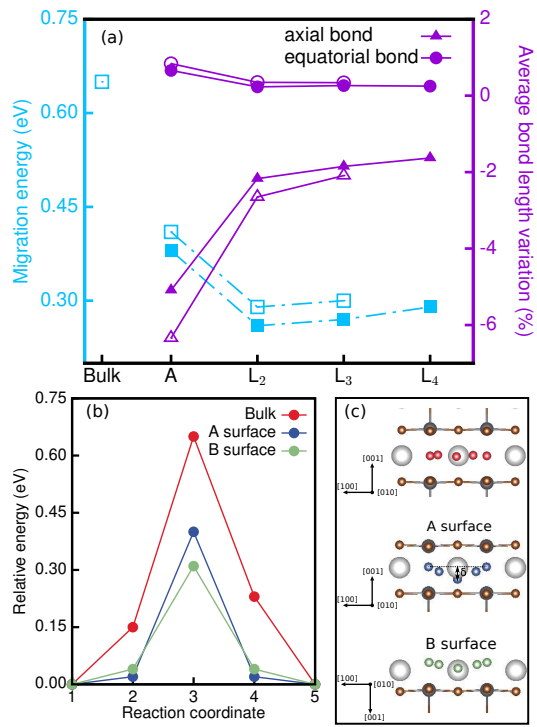}
    \caption{(a) Migration energy and average bond length variation in the slab structure as a function of the layer in which Br migration takes place. The migration energy in the bulk is also shown in blue for comparison. Open (closed) symbols correspond to the surface slab with three (four) mobile top layers. (b) Energy profiles and (c) migration paths, as computed with the cNEB approach for migration in the bulk (red), at the A surface (blue) and at the B surface (green). The structures correspond to the average atomic configurations of the two equivalent endpoints of each cNEB calculation overlaid with the position of the migrating Br ion along the migration path. The definition of the deviation $\delta$ from the linear path is shown in the structure corresponding to A surface.}
    \label{fig3:activation-energy-migration-path}
\end{figure}

Table~\ref{tbl2:activation_energy} and Figure~\ref{fig3:activation-energy-migration-path}(a) and (b) show the activation energy for Br migration at the A- and B-terminated surfaces, and within subsurface layers of surface A. Our first main finding is that the migration energy at both surfaces is substantially lower than that in the bulk, more than a factor of two at the B surface. We have confirmed that our finding of a significantly lower migration energy at the surface also holds in a $2\times2\times6$ unit cell setup, where we calculate migration energies of 0.48\,eV in the bulk and 0.28\,eV at the surface with values of 0.20\,eV and 0.26\,eV in surface layers L$_2$ and L$_3$). Furthermore we find that migration barriers at surface A and B are very similar; the migration barrier at surface A is only 90\,meV larger than that at surface B. Interestingly, the migration barrier in subsurface layer L$_2$ is $\sim$110 meV lower than directly at the surface, and only slowly increases in subsequent subsurface layers. We show in Figure~\ref{fig3:activation-energy-migration-path}(a) that the variation of the migration barrier with the layer number is correlated with relative bond lengths variations of axial and equatorial bond lengths with respect to the bulk. Smaller migration energies are associated with a significant axial compression of the surface. The slightly higher migration barrier at the surface of A as compared to subsequent surface layers is correlated with a subtle interplay between longer equatorial and shorter axial bond lengths as compared to the bulk. However, note that trends in migration barriers as a function of surface depth should be viewed with caution: In our calculations, the bottom layers of the surface slab are fixed to the bulk atomic positions, a constraint that might affect the magnitude of the calculated migration barriers in the layers adjacent to the fixed layers. We therefore calculated migration barriers for a surface slab with four mobile layers as well. The migration barriers for this system, also shown in Figure ~\ref{fig3:activation-energy-migration-path}(a) and Table~\ref{tbl2:activation_energy}, are slightly lower but follow the same trends. Observation of bulk-like migration barriers deeper into the surface would likely require structural models with more surface layers. Our results are also in line with observations by Oranskaia et al.~showing that larger lattice distortions lead to smaller migration energies for the through-cell migration of a Br vacancy in organic-inorganic perovskites \cite{Oranskaia2018}.

A detailed analysis of the migration paths of the Br ion, associated with the energy profiles shown in Figure~\ref{fig3:activation-energy-migration-path}(b), reveals significant qualitative differences between the migration paths in the bulk and at the two surfaces, see Figure~\ref{fig3:activation-energy-migration-path}(c). In the bulk, the halide ion moves along an almost straight line from one axial vacancy position to the other - the shortest possible path. Contrary to that, the migration path is curved at both surfaces, with the saddle point deviating from the straight line. Analogous curved paths for vacancy migration between an equatorial and an axial position have previously been reported for oxide perovskites based on neutron diffraction \cite{Malavasi2010,Yashima2003} and computational \cite{SaifulIslam2000, Munoz-Garcia2014} studies. More recently, similar curved paths have been reported for Pb-based halide perovskites as well \cite{Eames2015, Zhang2020}. We quantify the curvature of the migration path, $\delta$, by computing the perpendicular distance of the Br ion in the cNEB saddle point configuration to the straight line between the initial and final positions of the Br ion, schematically represented in Figure~\ref{fig3:activation-energy-migration-path}(c). As reported in Table~\ref{tbl2:activation_energy}, we find that in the bulk, the Br ion follows an almost straight line, with a deviation more than 7 times lower than at the A and B surfaces. This finding highlights the more flexible nature of the surface, which can deform and accommodate a defect more easily, explaining the lower energy of Br vacancy migration as compared to the bulk. Finally, we observe that at both surfaces, the saddle point is bowed away from the surface, with $\delta$ at the A surface almost double of what it is at the B surface. We find that $\delta$ is correlated with the compression of axial Pb-Br bond lengths and can be traced back to the structural symmetry of the two surfaces: Formation of a Br vacancy is associated with the breaking of one bond at the B surface, leading to less restructuring as compared to surface A, where two bonds are broken and both the Pb-Br layer above and below the vacancy adjust to vacancy formation and migration.

\subsection{Surface passivation with alkali halide monolayers}
Motivated by the correlation between migration barriers and surface restructuring, we investigate the effect of surface modification on Br vacancy migration energies. Surface modification is a common strategy for passivating surface and interfacial defect states in halide perovskites \cite{Xue2020}. Chemical surface treatment with organic ligands has been shown to increase photoluminescence lifetimes and quantum yields \cite{Dane2015, DeQuilettes2016}. However, organic ligands may lead to problems with stability. Therefore, alkali halides have recently been suggested as interface modifiers between the halide perovskite absorber and the electron- or hole-transport layers in solar cells, with some studies showing that they lead to enhanced stability and device performance \cite{Liu2018, Chen2019b}. Moreover, a first principles study by Apergi et al.~demonstrated that alkali halide surface modifiers allow for improved electronic level alignment between the halide perovskite absorber and NiO hole transport layer with wide tunability to match those of various perovskite compositions \cite{Apergi2020}.

\begin{figure}[ht]
    \centering
    \includegraphics[width=0.9\columnwidth]{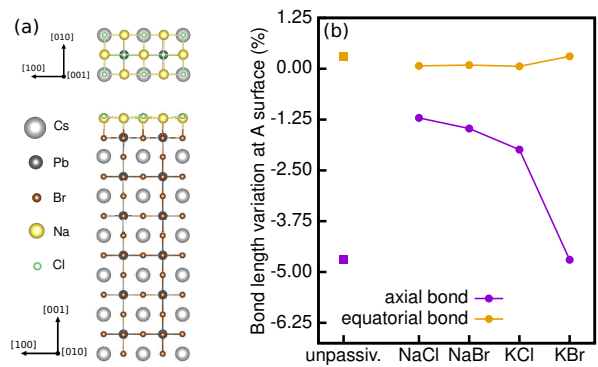}
    \caption{(a) Slab supercell passivated with NaCl monolayer at A-terminated surface (top and side views). (b) Pb-Br bond length variation at surface A of the slab structure as a function of passivation layer. For comparison, the Pb-Br bond lengths at surface A of the unpassivated slab structure are represented as squares.}
    \label{fig4:passivation}
\end{figure}

Here, we investigate four alkali halide monolayers, NaBr, NaCl, KBr and KCl, and their effect on Br vacancy migration at the surface of CsPbBr$_3$. We construct our passivated systems by placing the monolayer on top of surface A and relaxing the structures. In Figure~\ref{fig4:passivation}(a) we show the particular case of a slab structure passivated with a NaCl monolayer. Upon geometry optimization we find that the axial Pb-Br bonds of the surface slab are significantly less compressed than those of the unpassivated structure for NaCl- and NaBr-passivated surfaces. In Figure~\ref{fig4:passivation}(b), we show that the variation of Pb-Br axial bonds at the A-surface is less than 2\,\% and that of equatorial bonds is negligible. In comparison with the undistorted CsPbBr$_3$ bulk structure, K-based monolayers feature longer bonds, leading to larger distortions induced by the lattice mismatch between the perovskite and the passivation layer. In fact, passivating the surface with a KBr monolayer does not reduce distortions and yields very similar bonds as compared to the unpassivated system. In contrast, NaCl and NaBr monolayers have bond lengths that differ by only 0.08\,\AA~from the bulk. We therefore hypothesize that passivation with NaCl and NaBr should lead to an increase of the Br vacancy migration barrier at the surface.

Following the same approach as before, we introduce a Br vacancy in the first A-surface layer of the NaCl-passivated and NaBr-passivated surface slab structure, respectively, and calculate the energy of Br vacancy migration in the surface layer. For the NaCl-passivated system, we find a migration barrier of 0.57\,eV, only 80\,meV lower than that computed for migration in the bulk. Interestingly, however, the vacancy follows a curved migration path, with a larger deviation from the straight line ($\delta=1.68$\,\AA) than in the unpassivated system. Furthermore, we find that passivating the surface with NaBr leads to a migration barrier of 0.48\,eV, slightly larger than that computed for the migration at the surface of the unpassivated slab system. This result, reinforces our hypothesis that larger variations of Pb-Br bond lengths lead to smaller migration energies and indicates the potential of simple alkali halide salts for suppressing halogen vacancy migration at surfaces of halide perovskites.

\section{Conclusions}
In conclusion, we performed a first principles DFT study of Br vacancy migration in CsPbBr$_3$ and showed that the migration barrier within the close-packed bulk structure of cubic CsPbBr$_3$ is roughly twice as large as that at either of the distinctly terminated (001) surfaces of the system. Our calculations suggest that the significant reduction of the migration barrier at the surface is due to the "softer" structure of the surface which allows for significant bond lengths variations as compared to the bulk. Motivated by this observation, we studied the effect of surface modification with alkali halide monolayers and demonstrated that passivation with NaCl significantly decreases the structural distortions seen in the unpassivated surface, in particular the compression of the axial Pb-Br bonds. Consequently, NaCl passivation leads to an increase of the Br vacancy migration barrier at the surface back to almost the value it has in the bulk. Our results highlight the important role of surfaces in determining perovskite stability by facilitating ion migration. The dependence of vacancy activation barriers on Pb-Br bond lengths, in particular the importance of axial bond length compression, suggests that strain engineering, for example via epitaxial growth, could be another viable route for suppressing ion migration in halide perovskites \cite{Chen2020}. We believe that future computational studies should be directed towards elucidating the role of grain boundaries, in particular in polycrystalline MAPbI$_3$. With the advent of machine-learning force fields with DFT accuracy, reliable structural models of large supercells of organic-inorganic halide perovskites and the inclusion of temperature effects in large-scale molecular dynamics simulations have become computationally feasible \cite{Jinnouchi2019}.

\begin{acknowledgments}
We thank S. H\"uttner for valuable discussions. This work was supported by the Bavarian State Ministry of Science and the Arts through the Collaborative Research Network Solar Technologies go Hybrid (SolTech), the Elite Network Bavaria, and the German Research Foundation (DFG) through SFB840 B7, and through computational resources provided by the Bavarian Polymer Institute (BPI). R.-I. Biega acknowledges support by the DFG program GRK1640. 
\end{acknowledgments}


\begin{thebibliography}{62}%
\makeatletter
\providecommand \@ifxundefined [1]{%
 \@ifx{#1\undefined}
}%
\providecommand \@ifnum [1]{%
 \ifnum #1\expandafter \@firstoftwo
 \else \expandafter \@secondoftwo
 \fi
}%
\providecommand \@ifx [1]{%
 \ifx #1\expandafter \@firstoftwo
 \else \expandafter \@secondoftwo
 \fi
}%
\providecommand \natexlab [1]{#1}%
\providecommand \enquote  [1]{``#1''}%
\providecommand \bibnamefont  [1]{#1}%
\providecommand \bibfnamefont [1]{#1}%
\providecommand \citenamefont [1]{#1}%
\providecommand \href@noop [0]{\@secondoftwo}%
\providecommand \href [0]{\begingroup \@sanitize@url \@href}%
\providecommand \@href[1]{\@@startlink{#1}\@@href}%
\providecommand \@@href[1]{\endgroup#1\@@endlink}%
\providecommand \@sanitize@url [0]{\catcode `\\12\catcode `\$12\catcode
  `\&12\catcode `\#12\catcode `\^12\catcode `\_12\catcode `\%12\relax}%
\providecommand \@@startlink[1]{}%
\providecommand \@@endlink[0]{}%
\providecommand \url  [0]{\begingroup\@sanitize@url \@url }%
\providecommand \@url [1]{\endgroup\@href {#1}{\urlprefix }}%
\providecommand \urlprefix  [0]{URL }%
\providecommand \Eprint [0]{\href }%
\providecommand \doibase [0]{http://dx.doi.org/}%
\providecommand \selectlanguage [0]{\@gobble}%
\providecommand \bibinfo  [0]{\@secondoftwo}%
\providecommand \bibfield  [0]{\@secondoftwo}%
\providecommand \translation [1]{[#1]}%
\providecommand \BibitemOpen [0]{}%
\providecommand \bibitemStop [0]{}%
\providecommand \bibitemNoStop [0]{.\EOS\space}%
\providecommand \EOS [0]{\spacefactor3000\relax}%
\providecommand \BibitemShut  [1]{\csname bibitem#1\endcsname}%
\let\auto@bib@innerbib\@empty
\bibitem [{\citenamefont {Kojima}\ \emph {et~al.}(2009)\citenamefont {Kojima},
  \citenamefont {Teshima}, \citenamefont {Shirai},\ and\ \citenamefont
  {Miyasaka}}]{Kojima2009}%
  \BibitemOpen
  \bibfield  {author} {\bibinfo {author} {\bibfnamefont {A.}~\bibnamefont
  {Kojima}}, \bibinfo {author} {\bibfnamefont {K.}~\bibnamefont {Teshima}},
  \bibinfo {author} {\bibfnamefont {Y.}~\bibnamefont {Shirai}}, \ and\ \bibinfo
  {author} {\bibfnamefont {T.}~\bibnamefont {Miyasaka}},\ }\href {\doibase
  10.1021/ja809598r} {\bibfield  {journal} {\bibinfo  {journal} {J. Am. Chem.
  Soc.}\ }\textbf {\bibinfo {volume} {131}},\ \bibinfo {pages} {6050} (\bibinfo
  {year} {2009})}\BibitemShut {NoStop}%
\bibitem [{\citenamefont {Liu}\ \emph {et~al.}(2013)\citenamefont {Liu},
  \citenamefont {Johnston},\ and\ \citenamefont {Snaith}}]{Liu2013}%
  \BibitemOpen
  \bibfield  {author} {\bibinfo {author} {\bibfnamefont {M.}~\bibnamefont
  {Liu}}, \bibinfo {author} {\bibfnamefont {M.~B.}\ \bibnamefont {Johnston}}, \
  and\ \bibinfo {author} {\bibfnamefont {H.~J.}\ \bibnamefont {Snaith}},\
  }\href {\doibase 10.1038/nature12509} {\bibfield  {journal} {\bibinfo
  {journal} {Nature}\ }\textbf {\bibinfo {volume} {501}},\ \bibinfo {pages}
  {395} (\bibinfo {year} {2013})}\BibitemShut {NoStop}%
\bibitem [{\citenamefont {Stranks}\ and\ \citenamefont
  {Snaith}(2015)}]{Stranks2015b}%
  \BibitemOpen
  \bibfield  {author} {\bibinfo {author} {\bibfnamefont {S.~D.}\ \bibnamefont
  {Stranks}}\ and\ \bibinfo {author} {\bibfnamefont {H.~J.}\ \bibnamefont
  {Snaith}},\ }\href@noop {} {\bibfield  {journal} {\bibinfo  {journal} {Nat.
  Nanotechnol.}\ }\textbf {\bibinfo {volume} {10}},\ \bibinfo {pages} {391}
  (\bibinfo {year} {2015})}\BibitemShut {NoStop}%
\bibitem [{\citenamefont {Tan}\ \emph {et~al.}(2014)\citenamefont {Tan},
  \citenamefont {Moghaddam}, \citenamefont {Lai}, \citenamefont {Docampo},
  \citenamefont {Higler}, \citenamefont {Deschler}, \citenamefont {Price},
  \citenamefont {Sadhanala}, \citenamefont {Pazos}, \citenamefont
  {Credgington}, \citenamefont {Hanusch}, \citenamefont {Bein}, \citenamefont
  {Snaith},\ and\ \citenamefont {Friend}}]{Tan2014}%
  \BibitemOpen
  \bibfield  {author} {\bibinfo {author} {\bibfnamefont {Z.~K.}\ \bibnamefont
  {Tan}}, \bibinfo {author} {\bibfnamefont {R.~S.}\ \bibnamefont {Moghaddam}},
  \bibinfo {author} {\bibfnamefont {M.~L.}\ \bibnamefont {Lai}}, \bibinfo
  {author} {\bibfnamefont {P.}~\bibnamefont {Docampo}}, \bibinfo {author}
  {\bibfnamefont {R.}~\bibnamefont {Higler}}, \bibinfo {author} {\bibfnamefont
  {F.}~\bibnamefont {Deschler}}, \bibinfo {author} {\bibfnamefont
  {M.}~\bibnamefont {Price}}, \bibinfo {author} {\bibfnamefont
  {A.}~\bibnamefont {Sadhanala}}, \bibinfo {author} {\bibfnamefont {L.~M.}\
  \bibnamefont {Pazos}}, \bibinfo {author} {\bibfnamefont {D.}~\bibnamefont
  {Credgington}}, \bibinfo {author} {\bibfnamefont {F.}~\bibnamefont
  {Hanusch}}, \bibinfo {author} {\bibfnamefont {T.}~\bibnamefont {Bein}},
  \bibinfo {author} {\bibfnamefont {H.~J.}\ \bibnamefont {Snaith}}, \ and\
  \bibinfo {author} {\bibfnamefont {R.~H.}\ \bibnamefont {Friend}},\ }\href
  {\doibase 10.1038/nnano.2014.149} {\bibfield  {journal} {\bibinfo  {journal}
  {Nat. Nanotech.}\ }\textbf {\bibinfo {volume} {9}},\ \bibinfo {pages} {687}
  (\bibinfo {year} {2014})}\BibitemShut {NoStop}%
\bibitem [{\citenamefont {Ling}\ \emph {et~al.}(2016)\citenamefont {Ling},
  \citenamefont {Yuan}, \citenamefont {Tian}, \citenamefont {Wang},
  \citenamefont {Wang}, \citenamefont {Xin}, \citenamefont {Hanson},
  \citenamefont {Ma},\ and\ \citenamefont {Gao}}]{Ling2016}%
  \BibitemOpen
  \bibfield  {author} {\bibinfo {author} {\bibfnamefont {Y.}~\bibnamefont
  {Ling}}, \bibinfo {author} {\bibfnamefont {Z.}~\bibnamefont {Yuan}}, \bibinfo
  {author} {\bibfnamefont {Y.}~\bibnamefont {Tian}}, \bibinfo {author}
  {\bibfnamefont {X.}~\bibnamefont {Wang}}, \bibinfo {author} {\bibfnamefont
  {J.~C.}\ \bibnamefont {Wang}}, \bibinfo {author} {\bibfnamefont
  {Y.}~\bibnamefont {Xin}}, \bibinfo {author} {\bibfnamefont {K.}~\bibnamefont
  {Hanson}}, \bibinfo {author} {\bibfnamefont {B.}~\bibnamefont {Ma}}, \ and\
  \bibinfo {author} {\bibfnamefont {H.}~\bibnamefont {Gao}},\ }\href {\doibase
  10.1002/adma.201503954} {\bibfield  {journal} {\bibinfo  {journal} {Adv.
  Mater.}\ }\textbf {\bibinfo {volume} {28}},\ \bibinfo {pages} {305} (\bibinfo
  {year} {2016})}\BibitemShut {NoStop}%
\bibitem [{\citenamefont {Zhang}\ \emph {et~al.}(2018)\citenamefont {Zhang},
  \citenamefont {Zhang}, \citenamefont {Liu}, \citenamefont {Ju}, \citenamefont
  {Zhang}, \citenamefont {Cheng},\ and\ \citenamefont {Tao}}]{Zhang2018}%
  \BibitemOpen
  \bibfield  {author} {\bibinfo {author} {\bibfnamefont {P.}~\bibnamefont
  {Zhang}}, \bibinfo {author} {\bibfnamefont {G.}~\bibnamefont {Zhang}},
  \bibinfo {author} {\bibfnamefont {L.}~\bibnamefont {Liu}}, \bibinfo {author}
  {\bibfnamefont {D.}~\bibnamefont {Ju}}, \bibinfo {author} {\bibfnamefont
  {L.}~\bibnamefont {Zhang}}, \bibinfo {author} {\bibfnamefont
  {K.}~\bibnamefont {Cheng}}, \ and\ \bibinfo {author} {\bibfnamefont
  {X.}~\bibnamefont {Tao}},\ }\href {\doibase 10.1021/acs.jpclett.8b01945}
  {\bibfield  {journal} {\bibinfo  {journal} {J. Phys. Chem. Lett.}\ }\textbf
  {\bibinfo {volume} {9}},\ \bibinfo {pages} {5040} (\bibinfo {year}
  {2018})}\BibitemShut {NoStop}%
\bibitem [{\citenamefont {Saidaminov}\ \emph {et~al.}(2017)\citenamefont
  {Saidaminov}, \citenamefont {Haque}, \citenamefont {Almutlaq}, \citenamefont
  {Sarmah}, \citenamefont {Miao}, \citenamefont {Begum}, \citenamefont
  {Zhumekenov}, \citenamefont {Dursun}, \citenamefont {Cho}, \citenamefont
  {Murali}, \citenamefont {Mohammed}, \citenamefont {Wu},\ and\ \citenamefont
  {Bakr}}]{Saidaminov2017}%
  \BibitemOpen
  \bibfield  {author} {\bibinfo {author} {\bibfnamefont {M.~I.}\ \bibnamefont
  {Saidaminov}}, \bibinfo {author} {\bibfnamefont {M.~A.}\ \bibnamefont
  {Haque}}, \bibinfo {author} {\bibfnamefont {J.}~\bibnamefont {Almutlaq}},
  \bibinfo {author} {\bibfnamefont {S.}~\bibnamefont {Sarmah}}, \bibinfo
  {author} {\bibfnamefont {X.~H.}\ \bibnamefont {Miao}}, \bibinfo {author}
  {\bibfnamefont {R.}~\bibnamefont {Begum}}, \bibinfo {author} {\bibfnamefont
  {A.~A.}\ \bibnamefont {Zhumekenov}}, \bibinfo {author} {\bibfnamefont
  {I.}~\bibnamefont {Dursun}}, \bibinfo {author} {\bibfnamefont
  {N.}~\bibnamefont {Cho}}, \bibinfo {author} {\bibfnamefont {B.}~\bibnamefont
  {Murali}}, \bibinfo {author} {\bibfnamefont {O.~F.}\ \bibnamefont
  {Mohammed}}, \bibinfo {author} {\bibfnamefont {T.}~\bibnamefont {Wu}}, \ and\
  \bibinfo {author} {\bibfnamefont {O.~M.}\ \bibnamefont {Bakr}},\ }\href
  {\doibase 10.1002/adom.201600704} {\bibfield  {journal} {\bibinfo  {journal}
  {Adv. Opt. Mater.}\ }\textbf {\bibinfo {volume} {5}},\ \bibinfo {pages}
  {1600704} (\bibinfo {year} {2017})}\BibitemShut {NoStop}%
\bibitem [{\citenamefont {Ding}\ \emph {et~al.}(2017)\citenamefont {Ding},
  \citenamefont {Du}, \citenamefont {Zuo}, \citenamefont {Zhao}, \citenamefont
  {Cui},\ and\ \citenamefont {Zhan}}]{Ding2017}%
  \BibitemOpen
  \bibfield  {author} {\bibinfo {author} {\bibfnamefont {J.}~\bibnamefont
  {Ding}}, \bibinfo {author} {\bibfnamefont {S.}~\bibnamefont {Du}}, \bibinfo
  {author} {\bibfnamefont {Z.}~\bibnamefont {Zuo}}, \bibinfo {author}
  {\bibfnamefont {Y.}~\bibnamefont {Zhao}}, \bibinfo {author} {\bibfnamefont
  {H.}~\bibnamefont {Cui}}, \ and\ \bibinfo {author} {\bibfnamefont
  {X.}~\bibnamefont {Zhan}},\ }\href {\doibase 10.1021/acs.jpcc.7b01171}
  {\bibfield  {journal} {\bibinfo  {journal} {J. Phys. Chem. C}\ }\textbf
  {\bibinfo {volume} {121}},\ \bibinfo {pages} {4917} (\bibinfo {year}
  {2017})}\BibitemShut {NoStop}%
\bibitem [{\citenamefont {Zhou}\ \emph {et~al.}(2021)\citenamefont {Zhou},
  \citenamefont {Chen}, \citenamefont {Bakr},\ and\ \citenamefont
  {Mohammed}}]{Zhou2021}%
  \BibitemOpen
  \bibfield  {author} {\bibinfo {author} {\bibfnamefont {Y.}~\bibnamefont
  {Zhou}}, \bibinfo {author} {\bibfnamefont {J.}~\bibnamefont {Chen}}, \bibinfo
  {author} {\bibfnamefont {O.~M.}\ \bibnamefont {Bakr}}, \ and\ \bibinfo
  {author} {\bibfnamefont {O.~F.}\ \bibnamefont {Mohammed}},\ }\href {\doibase
  10.1021/acsenergylett.0c02430} {\bibfield  {journal} {\bibinfo  {journal}
  {ACS Energy Lett.}\ }\textbf {\bibinfo {volume} {6}},\ \bibinfo {pages} {739}
  (\bibinfo {year} {2021})}\BibitemShut {NoStop}%
\bibitem [{\citenamefont {{National Renewable Energy
  Laboratory}}(2021)}]{nrel2021}%
  \BibitemOpen
  \bibfield  {author} {\bibinfo {author} {\bibnamefont {{National Renewable
  Energy Laboratory}}},\ }\href
  {https://www.nrel.gov/pv/assets/pdfs/best-research-cell-efficiencies.20200104.pdf}
  {\enquote {\bibinfo {title} {{Best Research-Cell Efficiencies}},}\ }
  (\bibinfo {year} {2021})\BibitemShut {NoStop}%
\bibitem [{\citenamefont {Senocrate}\ \emph {et~al.}(2019)\citenamefont
  {Senocrate}, \citenamefont {Kim}, \citenamefont {Gr{\"{a}}tzel},\ and\
  \citenamefont {Maier}}]{Senocrate2019}%
  \BibitemOpen
  \bibfield  {author} {\bibinfo {author} {\bibfnamefont {A.}~\bibnamefont
  {Senocrate}}, \bibinfo {author} {\bibfnamefont {G.~Y.}\ \bibnamefont {Kim}},
  \bibinfo {author} {\bibfnamefont {M.}~\bibnamefont {Gr{\"{a}}tzel}}, \ and\
  \bibinfo {author} {\bibfnamefont {J.}~\bibnamefont {Maier}},\ }\href
  {\doibase 10.1021/acsenergylett.9b01605} {\bibfield  {journal} {\bibinfo
  {journal} {ACS Energy Lett.}\ }\textbf {\bibinfo {volume} {4}},\ \bibinfo
  {pages} {2859} (\bibinfo {year} {2019})}\BibitemShut {NoStop}%
\bibitem [{\citenamefont {Saliba}\ \emph {et~al.}(2016)\citenamefont {Saliba},
  \citenamefont {Matsui}, \citenamefont {Seo}, \citenamefont {Domanski},
  \citenamefont {Correa-Baena}, \citenamefont {Nazeeruddin}, \citenamefont
  {Zakeeruddin}, \citenamefont {Tress}, \citenamefont {Abate}, \citenamefont
  {Hagfeldt},\ and\ \citenamefont {Gr{\"{a}}tzel}}]{Saliba2016}%
  \BibitemOpen
  \bibfield  {author} {\bibinfo {author} {\bibfnamefont {M.}~\bibnamefont
  {Saliba}}, \bibinfo {author} {\bibfnamefont {T.}~\bibnamefont {Matsui}},
  \bibinfo {author} {\bibfnamefont {J.~Y.}\ \bibnamefont {Seo}}, \bibinfo
  {author} {\bibfnamefont {K.}~\bibnamefont {Domanski}}, \bibinfo {author}
  {\bibfnamefont {J.~P.}\ \bibnamefont {Correa-Baena}}, \bibinfo {author}
  {\bibfnamefont {M.~K.}\ \bibnamefont {Nazeeruddin}}, \bibinfo {author}
  {\bibfnamefont {S.~M.}\ \bibnamefont {Zakeeruddin}}, \bibinfo {author}
  {\bibfnamefont {W.}~\bibnamefont {Tress}}, \bibinfo {author} {\bibfnamefont
  {A.}~\bibnamefont {Abate}}, \bibinfo {author} {\bibfnamefont
  {A.}~\bibnamefont {Hagfeldt}}, \ and\ \bibinfo {author} {\bibfnamefont
  {M.}~\bibnamefont {Gr{\"{a}}tzel}},\ }\href {\doibase 10.1039/c5ee03874j}
  {\bibfield  {journal} {\bibinfo  {journal} {Energy Environ. Sci.}\ }\textbf
  {\bibinfo {volume} {9}},\ \bibinfo {pages} {1989} (\bibinfo {year}
  {2016})}\BibitemShut {NoStop}%
\bibitem [{\citenamefont {Schileo}\ and\ \citenamefont
  {Grancini}(2020)}]{Schileo2020}%
  \BibitemOpen
  \bibfield  {author} {\bibinfo {author} {\bibfnamefont {G.}~\bibnamefont
  {Schileo}}\ and\ \bibinfo {author} {\bibfnamefont {G.}~\bibnamefont
  {Grancini}},\ }\href {\doibase 10.1088/2515-7655/ab6cc4} {\bibfield
  {journal} {\bibinfo  {journal} {Journal of Physics: Energy}\ }\textbf
  {\bibinfo {volume} {2}},\ \bibinfo {pages} {021005} (\bibinfo {year}
  {2020})}\BibitemShut {NoStop}%
\bibitem [{\citenamefont {Huang}\ \emph {et~al.}(2016)\citenamefont {Huang},
  \citenamefont {Polavarapu}, \citenamefont {Sichert}, \citenamefont {Susha},
  \citenamefont {Urban},\ and\ \citenamefont {Rogach}}]{Huang2016a}%
  \BibitemOpen
  \bibfield  {author} {\bibinfo {author} {\bibfnamefont {H.}~\bibnamefont
  {Huang}}, \bibinfo {author} {\bibfnamefont {L.}~\bibnamefont {Polavarapu}},
  \bibinfo {author} {\bibfnamefont {J.~A.}\ \bibnamefont {Sichert}}, \bibinfo
  {author} {\bibfnamefont {A.~S.}\ \bibnamefont {Susha}}, \bibinfo {author}
  {\bibfnamefont {A.~S.}\ \bibnamefont {Urban}}, \ and\ \bibinfo {author}
  {\bibfnamefont {A.~L.}\ \bibnamefont {Rogach}},\ }\href {\doibase
  10.1038/am.2016.167} {\bibfield  {journal} {\bibinfo  {journal} {NPG Asia
  Mater.}\ }\textbf {\bibinfo {volume} {8}},\ \bibinfo {pages} {e328} (\bibinfo
  {year} {2016})}\BibitemShut {NoStop}%
\bibitem [{\citenamefont {He}\ \emph {et~al.}(2018)\citenamefont {He},
  \citenamefont {Matei}, \citenamefont {Jung}, \citenamefont {McCall},
  \citenamefont {Chen}, \citenamefont {Stoumpos}, \citenamefont {Liu},
  \citenamefont {Peters}, \citenamefont {Chung}, \citenamefont {Wessels},
  \citenamefont {Wasielewski}, \citenamefont {Dravid}, \citenamefont {Burger},\
  and\ \citenamefont {Kanatzidis}}]{He2018}%
  \BibitemOpen
  \bibfield  {author} {\bibinfo {author} {\bibfnamefont {Y.}~\bibnamefont
  {He}}, \bibinfo {author} {\bibfnamefont {L.}~\bibnamefont {Matei}}, \bibinfo
  {author} {\bibfnamefont {H.~J.}\ \bibnamefont {Jung}}, \bibinfo {author}
  {\bibfnamefont {K.~M.}\ \bibnamefont {McCall}}, \bibinfo {author}
  {\bibfnamefont {M.}~\bibnamefont {Chen}}, \bibinfo {author} {\bibfnamefont
  {C.~C.}\ \bibnamefont {Stoumpos}}, \bibinfo {author} {\bibfnamefont
  {Z.}~\bibnamefont {Liu}}, \bibinfo {author} {\bibfnamefont {J.~A.}\
  \bibnamefont {Peters}}, \bibinfo {author} {\bibfnamefont {D.~Y.}\
  \bibnamefont {Chung}}, \bibinfo {author} {\bibfnamefont {B.~W.}\ \bibnamefont
  {Wessels}}, \bibinfo {author} {\bibfnamefont {M.~R.}\ \bibnamefont
  {Wasielewski}}, \bibinfo {author} {\bibfnamefont {V.~P.}\ \bibnamefont
  {Dravid}}, \bibinfo {author} {\bibfnamefont {A.}~\bibnamefont {Burger}}, \
  and\ \bibinfo {author} {\bibfnamefont {M.~G.}\ \bibnamefont {Kanatzidis}},\
  }\href {\doibase 10.1038/s41467-018-04073-3} {\bibfield  {journal} {\bibinfo
  {journal} {Nat. Comm.}\ }\textbf {\bibinfo {volume} {9}},\ \bibinfo {pages}
  {1609} (\bibinfo {year} {2018})}\BibitemShut {NoStop}%
\bibitem [{\citenamefont {Zhang}\ \emph {et~al.}(2019)\citenamefont {Zhang},
  \citenamefont {Fu}, \citenamefont {Tang}, \citenamefont {Wang}, \citenamefont
  {Zhang}, \citenamefont {Yu}, \citenamefont {Wang}, \citenamefont {Zhang},\
  and\ \citenamefont {Xiao}}]{Zhang2019}%
  \BibitemOpen
  \bibfield  {author} {\bibinfo {author} {\bibfnamefont {H.}~\bibnamefont
  {Zhang}}, \bibinfo {author} {\bibfnamefont {X.}~\bibnamefont {Fu}}, \bibinfo
  {author} {\bibfnamefont {Y.}~\bibnamefont {Tang}}, \bibinfo {author}
  {\bibfnamefont {H.}~\bibnamefont {Wang}}, \bibinfo {author} {\bibfnamefont
  {C.}~\bibnamefont {Zhang}}, \bibinfo {author} {\bibfnamefont {W.~W.}\
  \bibnamefont {Yu}}, \bibinfo {author} {\bibfnamefont {X.}~\bibnamefont
  {Wang}}, \bibinfo {author} {\bibfnamefont {Y.}~\bibnamefont {Zhang}}, \ and\
  \bibinfo {author} {\bibfnamefont {M.}~\bibnamefont {Xiao}},\ }\href {\doibase
  10.1038/s41467-019-09047-7} {\bibfield  {journal} {\bibinfo  {journal} {Nat.
  Comm.}\ }\textbf {\bibinfo {volume} {10}},\ \bibinfo {pages} {1088} (\bibinfo
  {year} {2019})}\BibitemShut {NoStop}%
\bibitem [{\citenamefont {Mizusaki}\ \emph {et~al.}(1983)\citenamefont
  {Mizusaki}, \citenamefont {Arai},\ and\ \citenamefont
  {Fueki}}]{Mizusaki1983}%
  \BibitemOpen
  \bibfield  {author} {\bibinfo {author} {\bibfnamefont {J.}~\bibnamefont
  {Mizusaki}}, \bibinfo {author} {\bibfnamefont {K.}~\bibnamefont {Arai}}, \
  and\ \bibinfo {author} {\bibfnamefont {K.}~\bibnamefont {Fueki}},\ }\href
  {\doibase 10.1016/0167-2738(83)90025-5} {\bibfield  {journal} {\bibinfo
  {journal} {Solid State Ionics}\ }\textbf {\bibinfo {volume} {11}},\ \bibinfo
  {pages} {203} (\bibinfo {year} {1983})}\BibitemShut {NoStop}%
\bibitem [{\citenamefont {Mosconi}\ and\ \citenamefont {{De
  Angelis}}(2016)}]{Mosconi2016b}%
  \BibitemOpen
  \bibfield  {author} {\bibinfo {author} {\bibfnamefont {E.}~\bibnamefont
  {Mosconi}}\ and\ \bibinfo {author} {\bibfnamefont {F.}~\bibnamefont {{De
  Angelis}}},\ }\href {\doibase 10.1021/acsenergylett.6b00108} {\bibfield
  {journal} {\bibinfo  {journal} {ACS Energy Lett.}\ }\textbf {\bibinfo
  {volume} {1}},\ \bibinfo {pages} {182} (\bibinfo {year} {2016})}\BibitemShut
  {NoStop}%
\bibitem [{\citenamefont {Meloni}\ \emph {et~al.}(2016)\citenamefont {Meloni},
  \citenamefont {Moehl}, \citenamefont {Tress}, \citenamefont {Franckeviius},
  \citenamefont {Saliba}, \citenamefont {Lee}, \citenamefont {Gao},
  \citenamefont {Nazeeruddin}, \citenamefont {Zakeeruddin}, \citenamefont
  {Rothlisberger}, \citenamefont {Graetzel},\ and\ \citenamefont
  {Figures}}]{Meloni2016}%
  \BibitemOpen
  \bibfield  {author} {\bibinfo {author} {\bibfnamefont {S.}~\bibnamefont
  {Meloni}}, \bibinfo {author} {\bibfnamefont {T.}~\bibnamefont {Moehl}},
  \bibinfo {author} {\bibfnamefont {W.}~\bibnamefont {Tress}}, \bibinfo
  {author} {\bibfnamefont {M.}~\bibnamefont {Franckeviius}}, \bibinfo {author}
  {\bibfnamefont {M.}~\bibnamefont {Saliba}}, \bibinfo {author} {\bibfnamefont
  {Y.~H.}\ \bibnamefont {Lee}}, \bibinfo {author} {\bibfnamefont
  {P.}~\bibnamefont {Gao}}, \bibinfo {author} {\bibfnamefont {M.~K.}\
  \bibnamefont {Nazeeruddin}}, \bibinfo {author} {\bibfnamefont {S.~M.}\
  \bibnamefont {Zakeeruddin}}, \bibinfo {author} {\bibfnamefont
  {U.}~\bibnamefont {Rothlisberger}}, \bibinfo {author} {\bibfnamefont
  {M.}~\bibnamefont {Graetzel}}, \ and\ \bibinfo {author} {\bibfnamefont
  {S.}~\bibnamefont {Figures}},\ }\href {\doibase 10.1038/ncomms10334}
  {\bibfield  {journal} {\bibinfo  {journal} {Nat. Comm.}\ }\textbf {\bibinfo
  {volume} {7}},\ \bibinfo {pages} {10334} (\bibinfo {year}
  {2016})}\BibitemShut {NoStop}%
\bibitem [{\citenamefont {Luo}\ \emph {et~al.}(2017)\citenamefont {Luo},
  \citenamefont {Khoram}, \citenamefont {Brittman}, \citenamefont {Zhu},
  \citenamefont {Lai}, \citenamefont {Ong}, \citenamefont {Garnett},\ and\
  \citenamefont {Fenning}}]{Luo2017}%
  \BibitemOpen
  \bibfield  {author} {\bibinfo {author} {\bibfnamefont {Y.}~\bibnamefont
  {Luo}}, \bibinfo {author} {\bibfnamefont {P.}~\bibnamefont {Khoram}},
  \bibinfo {author} {\bibfnamefont {S.}~\bibnamefont {Brittman}}, \bibinfo
  {author} {\bibfnamefont {Z.}~\bibnamefont {Zhu}}, \bibinfo {author}
  {\bibfnamefont {B.}~\bibnamefont {Lai}}, \bibinfo {author} {\bibfnamefont
  {S.~P.}\ \bibnamefont {Ong}}, \bibinfo {author} {\bibfnamefont {E.~C.}\
  \bibnamefont {Garnett}}, \ and\ \bibinfo {author} {\bibfnamefont {D.~P.}\
  \bibnamefont {Fenning}},\ }\href {\doibase 10.1002/adma.201703451} {\bibfield
   {journal} {\bibinfo  {journal} {Adv. Mater.}\ }\textbf {\bibinfo {volume}
  {29}},\ \bibinfo {pages} {1703451} (\bibinfo {year} {2017})}\BibitemShut
  {NoStop}%
\bibitem [{\citenamefont {Senocrate}\ \emph {et~al.}(2017)\citenamefont
  {Senocrate}, \citenamefont {Moudrakovski}, \citenamefont {Kim}, \citenamefont
  {Yang}, \citenamefont {Gregori}, \citenamefont {Grätzel},\ and\
  \citenamefont {Maier}}]{Senocrate2017}%
  \BibitemOpen
  \bibfield  {author} {\bibinfo {author} {\bibfnamefont {A.}~\bibnamefont
  {Senocrate}}, \bibinfo {author} {\bibfnamefont {I.}~\bibnamefont
  {Moudrakovski}}, \bibinfo {author} {\bibfnamefont {G.~Y.}\ \bibnamefont
  {Kim}}, \bibinfo {author} {\bibfnamefont {T.-Y.}\ \bibnamefont {Yang}},
  \bibinfo {author} {\bibfnamefont {G.}~\bibnamefont {Gregori}}, \bibinfo
  {author} {\bibfnamefont {M.}~\bibnamefont {Grätzel}}, \ and\ \bibinfo
  {author} {\bibfnamefont {J.}~\bibnamefont {Maier}},\ }\href {\doibase
  10.1002/anie.201701724} {\bibfield  {journal} {\bibinfo  {journal} {Ang.
  Chem. Int. Ed.}\ }\textbf {\bibinfo {volume} {56}},\ \bibinfo {pages} {7755}
  (\bibinfo {year} {2017})}\BibitemShut {NoStop}%
\bibitem [{\citenamefont {Narayan}\ \emph {et~al.}(1987)\citenamefont
  {Narayan}, \citenamefont {Sarma},\ and\ \citenamefont
  {Suryanarayana}}]{Narayan1987}%
  \BibitemOpen
  \bibfield  {author} {\bibinfo {author} {\bibfnamefont {R.~L.}\ \bibnamefont
  {Narayan}}, \bibinfo {author} {\bibfnamefont {M.~V.}\ \bibnamefont {Sarma}},
  \ and\ \bibinfo {author} {\bibfnamefont {S.~V.}\ \bibnamefont
  {Suryanarayana}},\ }\href {\doibase 10.1007/BF01729441} {\bibfield  {journal}
  {\bibinfo  {journal} {J. Mater. Sci. Lett.}\ }\textbf {\bibinfo {volume}
  {6}},\ \bibinfo {pages} {93} (\bibinfo {year} {1987})}\BibitemShut {NoStop}%
\bibitem [{\citenamefont {Eames}\ \emph {et~al.}(2015)\citenamefont {Eames},
  \citenamefont {Frost}, \citenamefont {Barnes}, \citenamefont {O'Regan},
  \citenamefont {Walsh},\ and\ \citenamefont {Islam}}]{Eames2015}%
  \BibitemOpen
  \bibfield  {author} {\bibinfo {author} {\bibfnamefont {C.}~\bibnamefont
  {Eames}}, \bibinfo {author} {\bibfnamefont {J.~M.}\ \bibnamefont {Frost}},
  \bibinfo {author} {\bibfnamefont {P.~R.}\ \bibnamefont {Barnes}}, \bibinfo
  {author} {\bibfnamefont {B.~C.}\ \bibnamefont {O'Regan}}, \bibinfo {author}
  {\bibfnamefont {A.}~\bibnamefont {Walsh}}, \ and\ \bibinfo {author}
  {\bibfnamefont {M.~S.}\ \bibnamefont {Islam}},\ }\href {\doibase
  10.1038/ncomms8497} {\bibfield  {journal} {\bibinfo  {journal} {Nat. Comm.}\
  }\textbf {\bibinfo {volume} {6}},\ \bibinfo {pages} {7497} (\bibinfo {year}
  {2015})}\BibitemShut {NoStop}%
\bibitem [{\citenamefont {Yuan}\ and\ \citenamefont {Huang}(2016)}]{Yuan2016}%
  \BibitemOpen
  \bibfield  {author} {\bibinfo {author} {\bibfnamefont {Y.}~\bibnamefont
  {Yuan}}\ and\ \bibinfo {author} {\bibfnamefont {J.}~\bibnamefont {Huang}},\
  }\href {\doibase 10.1021/acs.accounts.5b00420} {\bibfield  {journal}
  {\bibinfo  {journal} {Acc. Chem. Res.}\ }\textbf {\bibinfo {volume} {49}},\
  \bibinfo {pages} {286} (\bibinfo {year} {2016})}\BibitemShut {NoStop}%
\bibitem [{\citenamefont {Yang}\ \emph {et~al.}(2015)\citenamefont {Yang},
  \citenamefont {Gregori}, \citenamefont {Pellet}, \citenamefont {Grätzel},\
  and\ \citenamefont {Maier}}]{Yang2015}%
  \BibitemOpen
  \bibfield  {author} {\bibinfo {author} {\bibfnamefont {T.-Y.}\ \bibnamefont
  {Yang}}, \bibinfo {author} {\bibfnamefont {G.}~\bibnamefont {Gregori}},
  \bibinfo {author} {\bibfnamefont {N.}~\bibnamefont {Pellet}}, \bibinfo
  {author} {\bibfnamefont {M.}~\bibnamefont {Grätzel}}, \ and\ \bibinfo
  {author} {\bibfnamefont {J.}~\bibnamefont {Maier}},\ }\href {\doibase
  10.1002/anie.201500014} {\bibfield  {journal} {\bibinfo  {journal} {Ang.
  Chem. Int. Ed.}\ }\textbf {\bibinfo {volume} {54}},\ \bibinfo {pages} {7905}
  (\bibinfo {year} {2015})}\BibitemShut {NoStop}%
\bibitem [{\citenamefont {Lee}\ \emph {et~al.}(2019)\citenamefont {Lee},
  \citenamefont {Kim}, \citenamefont {Yang}, \citenamefont {Yang},\ and\
  \citenamefont {Park}}]{Lee2019}%
  \BibitemOpen
  \bibfield  {author} {\bibinfo {author} {\bibfnamefont {J.~W.}\ \bibnamefont
  {Lee}}, \bibinfo {author} {\bibfnamefont {S.~G.}\ \bibnamefont {Kim}},
  \bibinfo {author} {\bibfnamefont {J.~M.}\ \bibnamefont {Yang}}, \bibinfo
  {author} {\bibfnamefont {Y.}~\bibnamefont {Yang}}, \ and\ \bibinfo {author}
  {\bibfnamefont {N.~G.}\ \bibnamefont {Park}},\ }\href {\doibase
  10.1063/1.5085643} {\bibfield  {journal} {\bibinfo  {journal} {APL Mater.}\
  }\textbf {\bibinfo {volume} {7}},\ \bibinfo {pages} {041111} (\bibinfo {year}
  {2019})}\BibitemShut {NoStop}%
\bibitem [{\citenamefont {Egger}\ \emph {et~al.}(2015)\citenamefont {Egger},
  \citenamefont {Kronik},\ and\ \citenamefont {Rappe}}]{Egger2015}%
  \BibitemOpen
  \bibfield  {author} {\bibinfo {author} {\bibfnamefont {D.~A.}\ \bibnamefont
  {Egger}}, \bibinfo {author} {\bibfnamefont {L.}~\bibnamefont {Kronik}}, \
  and\ \bibinfo {author} {\bibfnamefont {A.~M.}\ \bibnamefont {Rappe}},\ }\href
  {\doibase 10.1002/anie.201502544} {\bibfield  {journal} {\bibinfo  {journal}
  {Ang. Chem. Int. Ed.}\ }\textbf {\bibinfo {volume} {54}},\ \bibinfo {pages}
  {12437} (\bibinfo {year} {2015})}\BibitemShut {NoStop}%
\bibitem [{\citenamefont {Azpiroz}\ \emph {et~al.}(2015)\citenamefont
  {Azpiroz}, \citenamefont {Mosconi}, \citenamefont {Bisquert},\ and\
  \citenamefont {{De Angelis}}}]{Azpiroz2015}%
  \BibitemOpen
  \bibfield  {author} {\bibinfo {author} {\bibfnamefont {J.~M.}\ \bibnamefont
  {Azpiroz}}, \bibinfo {author} {\bibfnamefont {E.}~\bibnamefont {Mosconi}},
  \bibinfo {author} {\bibfnamefont {J.}~\bibnamefont {Bisquert}}, \ and\
  \bibinfo {author} {\bibfnamefont {F.}~\bibnamefont {{De Angelis}}},\ }\href
  {\doibase 10.1039/c5ee01265a} {\bibfield  {journal} {\bibinfo  {journal}
  {Energy Environ. Sci.}\ }\textbf {\bibinfo {volume} {8}},\ \bibinfo {pages}
  {2118} (\bibinfo {year} {2015})}\BibitemShut {NoStop}%
\bibitem [{\citenamefont {Haruyama}\ \emph {et~al.}(2015)\citenamefont
  {Haruyama}, \citenamefont {Sodeyama}, \citenamefont {Han},\ and\
  \citenamefont {Tateyama}}]{Haruyama2015}%
  \BibitemOpen
  \bibfield  {author} {\bibinfo {author} {\bibfnamefont {J.}~\bibnamefont
  {Haruyama}}, \bibinfo {author} {\bibfnamefont {K.}~\bibnamefont {Sodeyama}},
  \bibinfo {author} {\bibfnamefont {L.}~\bibnamefont {Han}}, \ and\ \bibinfo
  {author} {\bibfnamefont {Y.}~\bibnamefont {Tateyama}},\ }\href {\doibase
  10.1021/jacs.5b03615} {\bibfield  {journal} {\bibinfo  {journal} {J. Am.
  Chem. Soc.}\ }\textbf {\bibinfo {volume} {137}},\ \bibinfo {pages} {10048}
  (\bibinfo {year} {2015})}\BibitemShut {NoStop}%
\bibitem [{\citenamefont {Oranskaia}\ \emph {et~al.}(2018)\citenamefont
  {Oranskaia}, \citenamefont {Yin}, \citenamefont {Bakr}, \citenamefont
  {Br{\'{e}}das},\ and\ \citenamefont {Mohammed}}]{Oranskaia2018}%
  \BibitemOpen
  \bibfield  {author} {\bibinfo {author} {\bibfnamefont {A.}~\bibnamefont
  {Oranskaia}}, \bibinfo {author} {\bibfnamefont {J.}~\bibnamefont {Yin}},
  \bibinfo {author} {\bibfnamefont {O.~M.}\ \bibnamefont {Bakr}}, \bibinfo
  {author} {\bibfnamefont {J.~L.}\ \bibnamefont {Br{\'{e}}das}}, \ and\
  \bibinfo {author} {\bibfnamefont {O.~F.}\ \bibnamefont {Mohammed}},\ }\href
  {\doibase 10.1021/acs.jpclett.8b02522} {\bibfield  {journal} {\bibinfo
  {journal} {J. Phys. Chem. Lett.}\ }\textbf {\bibinfo {volume} {9}},\ \bibinfo
  {pages} {5474} (\bibinfo {year} {2018})}\BibitemShut {NoStop}%
\bibitem [{\citenamefont {Mosconi}\ \emph {et~al.}(2016)\citenamefont
  {Mosconi}, \citenamefont {Meggiolaro}, \citenamefont {Snaith}, \citenamefont
  {Stranks},\ and\ \citenamefont {De~Angelis}}]{Mosconi2016a}%
  \BibitemOpen
  \bibfield  {author} {\bibinfo {author} {\bibfnamefont {E.}~\bibnamefont
  {Mosconi}}, \bibinfo {author} {\bibfnamefont {D.}~\bibnamefont {Meggiolaro}},
  \bibinfo {author} {\bibfnamefont {H.~J.}\ \bibnamefont {Snaith}}, \bibinfo
  {author} {\bibfnamefont {S.~D.}\ \bibnamefont {Stranks}}, \ and\ \bibinfo
  {author} {\bibfnamefont {F.}~\bibnamefont {De~Angelis}},\ }\href {\doibase
  10.1039/C6EE01504B} {\bibfield  {journal} {\bibinfo  {journal} {Energy
  Environ. Sci.}\ }\textbf {\bibinfo {volume} {9}},\ \bibinfo {pages} {3180}
  (\bibinfo {year} {2016})}\BibitemShut {NoStop}%
\bibitem [{\citenamefont {Chen}\ \emph
  {et~al.}(2019{\natexlab{a}})\citenamefont {Chen}, \citenamefont {Fu},
  \citenamefont {Guo}, \citenamefont {Chen}, \citenamefont {Wang},
  \citenamefont {Luo},\ and\ \citenamefont {Zheng}}]{Chen2019a}%
  \BibitemOpen
  \bibfield  {author} {\bibinfo {author} {\bibfnamefont {C.}~\bibnamefont
  {Chen}}, \bibinfo {author} {\bibfnamefont {Q.}~\bibnamefont {Fu}}, \bibinfo
  {author} {\bibfnamefont {P.}~\bibnamefont {Guo}}, \bibinfo {author}
  {\bibfnamefont {H.}~\bibnamefont {Chen}}, \bibinfo {author} {\bibfnamefont
  {M.}~\bibnamefont {Wang}}, \bibinfo {author} {\bibfnamefont {W.}~\bibnamefont
  {Luo}}, \ and\ \bibinfo {author} {\bibfnamefont {Z.}~\bibnamefont {Zheng}},\
  }\href {\doibase 10.1088/2053-1591/ab4d79} {\bibfield  {journal} {\bibinfo
  {journal} {Mater. Res. Exp.}\ }\textbf {\bibinfo {volume} {6}},\ \bibinfo
  {pages} {115808} (\bibinfo {year} {2019}{\natexlab{a}})}\BibitemShut
  {NoStop}%
\bibitem [{\citenamefont {Futscher}\ \emph {et~al.}(2019)\citenamefont
  {Futscher}, \citenamefont {Lee}, \citenamefont {McGovern}, \citenamefont
  {Muscarella}, \citenamefont {Wang}, \citenamefont {Haider}, \citenamefont
  {Fakharuddin}, \citenamefont {Schmidt-Mende},\ and\ \citenamefont
  {Ehrler}}]{Futscher2019}%
  \BibitemOpen
  \bibfield  {author} {\bibinfo {author} {\bibfnamefont {M.~H.}\ \bibnamefont
  {Futscher}}, \bibinfo {author} {\bibfnamefont {J.~M.}\ \bibnamefont {Lee}},
  \bibinfo {author} {\bibfnamefont {L.}~\bibnamefont {McGovern}}, \bibinfo
  {author} {\bibfnamefont {L.~A.}\ \bibnamefont {Muscarella}}, \bibinfo
  {author} {\bibfnamefont {T.}~\bibnamefont {Wang}}, \bibinfo {author}
  {\bibfnamefont {M.~I.}\ \bibnamefont {Haider}}, \bibinfo {author}
  {\bibfnamefont {A.}~\bibnamefont {Fakharuddin}}, \bibinfo {author}
  {\bibfnamefont {L.}~\bibnamefont {Schmidt-Mende}}, \ and\ \bibinfo {author}
  {\bibfnamefont {B.}~\bibnamefont {Ehrler}},\ }\href {\doibase
  10.1039/c9mh00445a} {\bibfield  {journal} {\bibinfo  {journal} {Mater. Hor.}\
  }\textbf {\bibinfo {volume} {6}},\ \bibinfo {pages} {1497} (\bibinfo {year}
  {2019})}\BibitemShut {NoStop}%
\bibitem [{\citenamefont {Zhang}\ \emph
  {et~al.}(2020{\natexlab{a}})\citenamefont {Zhang}, \citenamefont {Hu},\ and\
  \citenamefont {Yang}}]{Zhang2020b}%
  \BibitemOpen
  \bibfield  {author} {\bibinfo {author} {\bibfnamefont {T.}~\bibnamefont
  {Zhang}}, \bibinfo {author} {\bibfnamefont {C.}~\bibnamefont {Hu}}, \ and\
  \bibinfo {author} {\bibfnamefont {S.}~\bibnamefont {Yang}},\ }\href@noop {}
  {\bibfield  {journal} {\bibinfo  {journal} {Small Methods}\ }\textbf
  {\bibinfo {volume} {4}},\ \bibinfo {pages} {1900552} (\bibinfo {year}
  {2020}{\natexlab{a}})}\BibitemShut {NoStop}%
\bibitem [{\citenamefont {Xing}\ \emph {et~al.}(2016)\citenamefont {Xing},
  \citenamefont {Wang}, \citenamefont {Dong}, \citenamefont {Yuan},
  \citenamefont {Fang},\ and\ \citenamefont {Huang}}]{Xing2016}%
  \BibitemOpen
  \bibfield  {author} {\bibinfo {author} {\bibfnamefont {J.}~\bibnamefont
  {Xing}}, \bibinfo {author} {\bibfnamefont {Q.}~\bibnamefont {Wang}}, \bibinfo
  {author} {\bibfnamefont {Q.}~\bibnamefont {Dong}}, \bibinfo {author}
  {\bibfnamefont {Y.}~\bibnamefont {Yuan}}, \bibinfo {author} {\bibfnamefont
  {Y.}~\bibnamefont {Fang}}, \ and\ \bibinfo {author} {\bibfnamefont
  {J.}~\bibnamefont {Huang}},\ }\href {\doibase 10.1039/c6cp06496e} {\bibfield
  {journal} {\bibinfo  {journal} {Phys. Chem. Chem. Phys.}\ }\textbf {\bibinfo
  {volume} {18}},\ \bibinfo {pages} {30484} (\bibinfo {year}
  {2016})}\BibitemShut {NoStop}%
\bibitem [{\citenamefont {Yun}\ \emph {et~al.}(2016)\citenamefont {Yun},
  \citenamefont {Seidel}, \citenamefont {Kim}, \citenamefont {Soufiani},
  \citenamefont {Huang}, \citenamefont {Lau}, \citenamefont {Jeon},
  \citenamefont {Seok}, \citenamefont {Green},\ and\ \citenamefont
  {Ho-Baillie}}]{Yun2016}%
  \BibitemOpen
  \bibfield  {author} {\bibinfo {author} {\bibfnamefont {J.~S.}\ \bibnamefont
  {Yun}}, \bibinfo {author} {\bibfnamefont {J.}~\bibnamefont {Seidel}},
  \bibinfo {author} {\bibfnamefont {J.}~\bibnamefont {Kim}}, \bibinfo {author}
  {\bibfnamefont {A.~M.}\ \bibnamefont {Soufiani}}, \bibinfo {author}
  {\bibfnamefont {S.}~\bibnamefont {Huang}}, \bibinfo {author} {\bibfnamefont
  {J.}~\bibnamefont {Lau}}, \bibinfo {author} {\bibfnamefont {N.~J.}\
  \bibnamefont {Jeon}}, \bibinfo {author} {\bibfnamefont {S.~I.}\ \bibnamefont
  {Seok}}, \bibinfo {author} {\bibfnamefont {M.~A.}\ \bibnamefont {Green}}, \
  and\ \bibinfo {author} {\bibfnamefont {A.}~\bibnamefont {Ho-Baillie}},\
  }\href {\doibase 10.1002/aenm.201600330} {\bibfield  {journal} {\bibinfo
  {journal} {Adv. Energy Mater.}\ }\textbf {\bibinfo {volume} {6}},\ \bibinfo
  {pages} {1600330} (\bibinfo {year} {2016})}\BibitemShut {NoStop}%
\bibitem [{\citenamefont {Meggiolaro}\ \emph {et~al.}(2019)\citenamefont
  {Meggiolaro}, \citenamefont {Mosconi},\ and\ \citenamefont {{De
  Angelis}}}]{Meggiolaro2019}%
  \BibitemOpen
  \bibfield  {author} {\bibinfo {author} {\bibfnamefont {D.}~\bibnamefont
  {Meggiolaro}}, \bibinfo {author} {\bibfnamefont {E.}~\bibnamefont {Mosconi}},
  \ and\ \bibinfo {author} {\bibfnamefont {F.}~\bibnamefont {{De Angelis}}},\
  }\href {\doibase 10.1021/acsenergylett.9b00247} {\bibfield  {journal}
  {\bibinfo  {journal} {ACS Energy Lett.}\ }\textbf {\bibinfo {volume} {4}},\
  \bibinfo {pages} {779} (\bibinfo {year} {2019})}\BibitemShut {NoStop}%
\bibitem [{\citenamefont {Quarti}\ \emph {et~al.}(2014)\citenamefont {Quarti},
  \citenamefont {Mosconi}, \citenamefont {{De Angelis}},\ and\ \citenamefont
  {Angelis}}]{Quarti2014}%
  \BibitemOpen
  \bibfield  {author} {\bibinfo {author} {\bibfnamefont {C.}~\bibnamefont
  {Quarti}}, \bibinfo {author} {\bibfnamefont {E.}~\bibnamefont {Mosconi}},
  \bibinfo {author} {\bibfnamefont {F.}~\bibnamefont {{De Angelis}}}, \ and\
  \bibinfo {author} {\bibfnamefont {F.~D.}\ \bibnamefont {Angelis}},\
  }\href@noop {} {\bibfield  {journal} {\bibinfo  {journal} {Chem. Mater.}\
  }\textbf {\bibinfo {volume} {26}},\ \bibinfo {pages} {6557} (\bibinfo {year}
  {2014})}\BibitemShut {NoStop}%
\bibitem [{\citenamefont {Hirotsu}\ \emph {et~al.}(1974)\citenamefont
  {Hirotsu}, \citenamefont {Harada}, \citenamefont {Iizumi},\ and\
  \citenamefont {Gesi}}]{Hirotsu1974}%
  \BibitemOpen
  \bibfield  {author} {\bibinfo {author} {\bibfnamefont {S.}~\bibnamefont
  {Hirotsu}}, \bibinfo {author} {\bibfnamefont {J.}~\bibnamefont {Harada}},
  \bibinfo {author} {\bibfnamefont {M.}~\bibnamefont {Iizumi}}, \ and\ \bibinfo
  {author} {\bibfnamefont {K.}~\bibnamefont {Gesi}},\ }\href@noop {} {\bibfield
   {journal} {\bibinfo  {journal} {J. Phys. Soc. Jap.}\ }\textbf {\bibinfo
  {volume} {37}},\ \bibinfo {pages} {1393} (\bibinfo {year}
  {1974})}\BibitemShut {NoStop}%
\bibitem [{\citenamefont {Stoumpos}\ \emph {et~al.}(2013)\citenamefont
  {Stoumpos}, \citenamefont {Malliakas}, \citenamefont {Peters}, \citenamefont
  {Liu}, \citenamefont {Sebastian}, \citenamefont {Im}, \citenamefont
  {Chasapis}, \citenamefont {Wibowo}, \citenamefont {Chung}, \citenamefont
  {Freeman}, \citenamefont {Wessels},\ and\ \citenamefont
  {Kanatzidis}}]{Stoumpos2013}%
  \BibitemOpen
  \bibfield  {author} {\bibinfo {author} {\bibfnamefont {C.~C.}\ \bibnamefont
  {Stoumpos}}, \bibinfo {author} {\bibfnamefont {C.~D.}\ \bibnamefont
  {Malliakas}}, \bibinfo {author} {\bibfnamefont {J.~A.}\ \bibnamefont
  {Peters}}, \bibinfo {author} {\bibfnamefont {Z.}~\bibnamefont {Liu}},
  \bibinfo {author} {\bibfnamefont {M.}~\bibnamefont {Sebastian}}, \bibinfo
  {author} {\bibfnamefont {J.}~\bibnamefont {Im}}, \bibinfo {author}
  {\bibfnamefont {T.~C.}\ \bibnamefont {Chasapis}}, \bibinfo {author}
  {\bibfnamefont {A.~C.}\ \bibnamefont {Wibowo}}, \bibinfo {author}
  {\bibfnamefont {D.~Y.}\ \bibnamefont {Chung}}, \bibinfo {author}
  {\bibfnamefont {A.~J.}\ \bibnamefont {Freeman}}, \bibinfo {author}
  {\bibfnamefont {B.~W.}\ \bibnamefont {Wessels}}, \ and\ \bibinfo {author}
  {\bibfnamefont {M.~G.}\ \bibnamefont {Kanatzidis}},\ }\href {\doibase
  10.1021/cg400645t} {\bibfield  {journal} {\bibinfo  {journal} {Crystal Growth
  and Design}\ }\textbf {\bibinfo {volume} {13}},\ \bibinfo {pages} {2722}
  (\bibinfo {year} {2013})}\BibitemShut {NoStop}%
\bibitem [{\citenamefont {Perdew}\ \emph {et~al.}(2008)\citenamefont {Perdew},
  \citenamefont {Ruzsinszky}, \citenamefont {Csonka}, \citenamefont {Vydrov},
  \citenamefont {Scuseria}, \citenamefont {Constantin}, \citenamefont {Zhou},\
  and\ \citenamefont {Burke}}]{Perdew2008}%
  \BibitemOpen
  \bibfield  {author} {\bibinfo {author} {\bibfnamefont {J.~P.}\ \bibnamefont
  {Perdew}}, \bibinfo {author} {\bibfnamefont {A.}~\bibnamefont {Ruzsinszky}},
  \bibinfo {author} {\bibfnamefont {G.~I.}\ \bibnamefont {Csonka}}, \bibinfo
  {author} {\bibfnamefont {O.~A.}\ \bibnamefont {Vydrov}}, \bibinfo {author}
  {\bibfnamefont {G.~E.}\ \bibnamefont {Scuseria}}, \bibinfo {author}
  {\bibfnamefont {L.~A.}\ \bibnamefont {Constantin}}, \bibinfo {author}
  {\bibfnamefont {X.}~\bibnamefont {Zhou}}, \ and\ \bibinfo {author}
  {\bibfnamefont {K.}~\bibnamefont {Burke}},\ }\href {\doibase
  10.1103/PhysRevLett.100.136406} {\bibfield  {journal} {\bibinfo  {journal}
  {Phys. Rev. Lett.}\ }\textbf {\bibinfo {volume} {100}},\ \bibinfo {pages}
  {136406} (\bibinfo {year} {2008})}\BibitemShut {NoStop}%
\bibitem [{\citenamefont {Kresse}\ and\ \citenamefont
  {Hafner}(1993)}]{Kresse1993}%
  \BibitemOpen
  \bibfield  {author} {\bibinfo {author} {\bibfnamefont {G.}~\bibnamefont
  {Kresse}}\ and\ \bibinfo {author} {\bibfnamefont {J.}~\bibnamefont
  {Hafner}},\ }\href {\doibase 10.1103/PhysRevB.47.558} {\bibfield  {journal}
  {\bibinfo  {journal} {Phys. Rev. B}\ }\textbf {\bibinfo {volume} {47}},\
  \bibinfo {pages} {558} (\bibinfo {year} {1993})}\BibitemShut {NoStop}%
\bibitem [{\citenamefont {Kresse}\ and\ \citenamefont
  {Furthm\"uller}(1996)}]{Kresse1996}%
  \BibitemOpen
  \bibfield  {author} {\bibinfo {author} {\bibfnamefont {G.}~\bibnamefont
  {Kresse}}\ and\ \bibinfo {author} {\bibfnamefont {J.}~\bibnamefont
  {Furthm\"uller}},\ }\href {\doibase 10.1103/PhysRevB.54.11169} {\bibfield
  {journal} {\bibinfo  {journal} {Phys. Rev. B}\ }\textbf {\bibinfo {volume}
  {54}},\ \bibinfo {pages} {11169} (\bibinfo {year} {1996})}\BibitemShut
  {NoStop}%
\bibitem [{\citenamefont {L{\'{o}}pez}\ \emph {et~al.}(2020)\citenamefont
  {L{\'{o}}pez}, \citenamefont {Abia}, \citenamefont {Alvarez-Galv{\'{a}}n},
  \citenamefont {Hong}, \citenamefont {Mart{\'{i}}nez-Huerta}, \citenamefont
  {Serrano-S{\'{a}}nchez}, \citenamefont {Carrascoso}, \citenamefont
  {Castellanos-G{\'{o}}mez}, \citenamefont {Fern{\'{a}}ndez-Dĺaz},\ and\
  \citenamefont {Alonso}}]{Lopez2020}%
  \BibitemOpen
  \bibfield  {author} {\bibinfo {author} {\bibfnamefont {C.~A.}\ \bibnamefont
  {L{\'{o}}pez}}, \bibinfo {author} {\bibfnamefont {C.}~\bibnamefont {Abia}},
  \bibinfo {author} {\bibfnamefont {M.~C.}\ \bibnamefont
  {Alvarez-Galv{\'{a}}n}}, \bibinfo {author} {\bibfnamefont {B.~K.}\
  \bibnamefont {Hong}}, \bibinfo {author} {\bibfnamefont {M.~V.}\ \bibnamefont
  {Mart{\'{i}}nez-Huerta}}, \bibinfo {author} {\bibfnamefont {F.}~\bibnamefont
  {Serrano-S{\'{a}}nchez}}, \bibinfo {author} {\bibfnamefont {F.}~\bibnamefont
  {Carrascoso}}, \bibinfo {author} {\bibfnamefont {A.}~\bibnamefont
  {Castellanos-G{\'{o}}mez}}, \bibinfo {author} {\bibfnamefont {M.~T.}\
  \bibnamefont {Fern{\'{a}}ndez-Dĺaz}}, \ and\ \bibinfo {author}
  {\bibfnamefont {J.~A.}\ \bibnamefont {Alonso}},\ }\href {\doibase
  10.1021/acsomega.9b04248} {\bibfield  {journal} {\bibinfo  {journal} {ACS
  Omega}\ }\textbf {\bibinfo {volume} {5}},\ \bibinfo {pages} {5931} (\bibinfo
  {year} {2020})}\BibitemShut {NoStop}%
\bibitem [{\citenamefont {Klime\ifmmode~\check{s}\else \v{s}\fi{}}\ \emph
  {et~al.}(2014)\citenamefont {Klime\ifmmode~\check{s}\else \v{s}\fi{}},
  \citenamefont {Kaltak},\ and\ \citenamefont {Kresse}}]{Kresse2014}%
  \BibitemOpen
  \bibfield  {author} {\bibinfo {author} {\bibfnamefont {J.~c.~v.}\
  \bibnamefont {Klime\ifmmode~\check{s}\else \v{s}\fi{}}}, \bibinfo {author}
  {\bibfnamefont {M.}~\bibnamefont {Kaltak}}, \ and\ \bibinfo {author}
  {\bibfnamefont {G.}~\bibnamefont {Kresse}},\ }\href {\doibase
  10.1103/PhysRevB.90.075125} {\bibfield  {journal} {\bibinfo  {journal} {Phys.
  Rev. B}\ }\textbf {\bibinfo {volume} {90}},\ \bibinfo {pages} {075125}
  (\bibinfo {year} {2014})}\BibitemShut {NoStop}%
\bibitem [{\citenamefont {Henkelman}\ and\ \citenamefont
  {Jónsson}(2000)}]{Henkelman2000}%
  \BibitemOpen
  \bibfield  {author} {\bibinfo {author} {\bibfnamefont {G.}~\bibnamefont
  {Henkelman}}\ and\ \bibinfo {author} {\bibfnamefont {H.}~\bibnamefont
  {Jónsson}},\ }\href@noop {} {\bibfield  {journal} {\bibinfo  {journal} {J.
  Chem. Phys.}\ }\textbf {\bibinfo {volume} {113}},\ \bibinfo {pages} {9901}
  (\bibinfo {year} {2000})}\BibitemShut {NoStop}%
\bibitem [{\citenamefont {Kang}\ and\ \citenamefont {Wang}(2017)}]{Kang2017}%
  \BibitemOpen
  \bibfield  {author} {\bibinfo {author} {\bibfnamefont {J.}~\bibnamefont
  {Kang}}\ and\ \bibinfo {author} {\bibfnamefont {L.-W.}\ \bibnamefont
  {Wang}},\ }\href {\doibase 10.1021/acs.jpclett.6b02800} {\bibfield  {journal}
  {\bibinfo  {journal} {J. Phys. Chem. Lett.}\ }\textbf {\bibinfo {volume}
  {8}},\ \bibinfo {pages} {489} (\bibinfo {year} {2017})}\BibitemShut {NoStop}%
\bibitem [{\citenamefont {Zhang}\ and\ \citenamefont {Sit}(2019)}]{Zhang2019a}%
  \BibitemOpen
  \bibfield  {author} {\bibinfo {author} {\bibfnamefont {L.}~\bibnamefont
  {Zhang}}\ and\ \bibinfo {author} {\bibfnamefont {P.~H.-L.}\ \bibnamefont
  {Sit}},\ }\href {\doibase 10.1039/C8TA09512D} {\bibfield  {journal} {\bibinfo
   {journal} {J. Mater. Chem. A}\ }\textbf {\bibinfo {volume} {7}},\ \bibinfo
  {pages} {2135} (\bibinfo {year} {2019})}\BibitemShut {NoStop}%
\bibitem [{\citenamefont {Zhang}\ \emph
  {et~al.}(2020{\natexlab{b}})\citenamefont {Zhang}, \citenamefont {Wang},
  \citenamefont {Zhang}, \citenamefont {Xiao}, \citenamefont {Sun},
  \citenamefont {Guo}, \citenamefont {Hafsia}, \citenamefont {Shao},
  \citenamefont {Xu},\ and\ \citenamefont {Zhou}}]{Zhang2020}%
  \BibitemOpen
  \bibfield  {author} {\bibinfo {author} {\bibfnamefont {B.-b.~B.}\
  \bibnamefont {Zhang}}, \bibinfo {author} {\bibfnamefont {F.}~\bibnamefont
  {Wang}}, \bibinfo {author} {\bibfnamefont {H.}~\bibnamefont {Zhang}},
  \bibinfo {author} {\bibfnamefont {B.}~\bibnamefont {Xiao}}, \bibinfo {author}
  {\bibfnamefont {Q.}~\bibnamefont {Sun}}, \bibinfo {author} {\bibfnamefont
  {J.}~\bibnamefont {Guo}}, \bibinfo {author} {\bibfnamefont {A.~B.}\
  \bibnamefont {Hafsia}}, \bibinfo {author} {\bibfnamefont {A.}~\bibnamefont
  {Shao}}, \bibinfo {author} {\bibfnamefont {Y.}~\bibnamefont {Xu}}, \ and\
  \bibinfo {author} {\bibfnamefont {J.}~\bibnamefont {Zhou}},\ }\href {\doibase
  10.1063/1.5134108} {\bibfield  {journal} {\bibinfo  {journal} {Appl. Phys.
  Lett.}\ }\textbf {\bibinfo {volume} {116}},\ \bibinfo {pages} {063505}
  (\bibinfo {year} {2020}{\natexlab{b}})}\BibitemShut {NoStop}%
\bibitem [{\citenamefont {Yaffe}\ \emph {et~al.}(2017)\citenamefont {Yaffe},
  \citenamefont {Guo}, \citenamefont {Tan}, \citenamefont {Egger},
  \citenamefont {Hull}, \citenamefont {Stoumpos}, \citenamefont {Zheng},
  \citenamefont {Heinz}, \citenamefont {Kronik}, \citenamefont {Kanatzidis},
  \citenamefont {Owen}, \citenamefont {Rappe}, \citenamefont {Pimenta},\ and\
  \citenamefont {Brus}}]{Yaffe2017}%
  \BibitemOpen
  \bibfield  {author} {\bibinfo {author} {\bibfnamefont {O.}~\bibnamefont
  {Yaffe}}, \bibinfo {author} {\bibfnamefont {Y.}~\bibnamefont {Guo}}, \bibinfo
  {author} {\bibfnamefont {L.~Z.}\ \bibnamefont {Tan}}, \bibinfo {author}
  {\bibfnamefont {D.~A.}\ \bibnamefont {Egger}}, \bibinfo {author}
  {\bibfnamefont {T.}~\bibnamefont {Hull}}, \bibinfo {author} {\bibfnamefont
  {C.~C.}\ \bibnamefont {Stoumpos}}, \bibinfo {author} {\bibfnamefont
  {F.}~\bibnamefont {Zheng}}, \bibinfo {author} {\bibfnamefont {T.~F.}\
  \bibnamefont {Heinz}}, \bibinfo {author} {\bibfnamefont {L.}~\bibnamefont
  {Kronik}}, \bibinfo {author} {\bibfnamefont {M.~G.}\ \bibnamefont
  {Kanatzidis}}, \bibinfo {author} {\bibfnamefont {J.~S.}\ \bibnamefont
  {Owen}}, \bibinfo {author} {\bibfnamefont {A.~M.}\ \bibnamefont {Rappe}},
  \bibinfo {author} {\bibfnamefont {M.~A.}\ \bibnamefont {Pimenta}}, \ and\
  \bibinfo {author} {\bibfnamefont {L.~E.}\ \bibnamefont {Brus}},\ }\href
  {\doibase 10.1103/PhysRevLett.118.136001} {\bibfield  {journal} {\bibinfo
  {journal} {Phys. Rev. Lett.}\ }\textbf {\bibinfo {volume} {118}},\ \bibinfo
  {pages} {136001} (\bibinfo {year} {2017})}\BibitemShut {NoStop}%
\bibitem [{\citenamefont {Malavasi}\ \emph {et~al.}(2010)\citenamefont
  {Malavasi}, \citenamefont {Fisher},\ and\ \citenamefont
  {Islam}}]{Malavasi2010}%
  \BibitemOpen
  \bibfield  {author} {\bibinfo {author} {\bibfnamefont {L.}~\bibnamefont
  {Malavasi}}, \bibinfo {author} {\bibfnamefont {C.~A.~J.}\ \bibnamefont
  {Fisher}}, \ and\ \bibinfo {author} {\bibfnamefont {M.~S.}\ \bibnamefont
  {Islam}},\ }\href {\doibase 10.1039/B915141A} {\bibfield  {journal} {\bibinfo
   {journal} {Chem. Soc. Rev.}\ }\textbf {\bibinfo {volume} {39}},\ \bibinfo
  {pages} {4370} (\bibinfo {year} {2010})}\BibitemShut {NoStop}%
\bibitem [{\citenamefont {Yashima}\ \emph {et~al.}(2003)\citenamefont
  {Yashima}, \citenamefont {Nomura}, \citenamefont {Kageyama}, \citenamefont
  {Miyazaki}, \citenamefont {Chitose},\ and\ \citenamefont
  {Adachi}}]{Yashima2003}%
  \BibitemOpen
  \bibfield  {author} {\bibinfo {author} {\bibfnamefont {M.}~\bibnamefont
  {Yashima}}, \bibinfo {author} {\bibfnamefont {K.}~\bibnamefont {Nomura}},
  \bibinfo {author} {\bibfnamefont {H.}~\bibnamefont {Kageyama}}, \bibinfo
  {author} {\bibfnamefont {Y.}~\bibnamefont {Miyazaki}}, \bibinfo {author}
  {\bibfnamefont {N.}~\bibnamefont {Chitose}}, \ and\ \bibinfo {author}
  {\bibfnamefont {K.}~\bibnamefont {Adachi}},\ }\href {\doibase
  https://doi.org/10.1016/j.cplett.2003.08.121} {\bibfield  {journal} {\bibinfo
   {journal} {Chem. Phys. Lett.}\ }\textbf {\bibinfo {volume} {380}},\ \bibinfo
  {pages} {391} (\bibinfo {year} {2003})}\BibitemShut {NoStop}%
\bibitem [{\citenamefont {Saiful~Islam}(2000)}]{SaifulIslam2000}%
  \BibitemOpen
  \bibfield  {author} {\bibinfo {author} {\bibfnamefont {M.}~\bibnamefont
  {Saiful~Islam}},\ }\href {\doibase 10.1039/A908425H} {\bibfield  {journal}
  {\bibinfo  {journal} {J. Mater. Chem.}\ }\textbf {\bibinfo {volume} {10}},\
  \bibinfo {pages} {1027} (\bibinfo {year} {2000})}\BibitemShut {NoStop}%
\bibitem [{\citenamefont {Mu{\~{n}}oz-Garc{\'{i}}a}\ \emph
  {et~al.}(2014)\citenamefont {Mu{\~{n}}oz-Garc{\'{i}}a}, \citenamefont
  {Ritzmann}, \citenamefont {Pavone}, \citenamefont {Keith},\ and\
  \citenamefont {Carter}}]{Munoz-Garcia2014}%
  \BibitemOpen
  \bibfield  {author} {\bibinfo {author} {\bibfnamefont {A.~B.}\ \bibnamefont
  {Mu{\~{n}}oz-Garc{\'{i}}a}}, \bibinfo {author} {\bibfnamefont {A.~M.}\
  \bibnamefont {Ritzmann}}, \bibinfo {author} {\bibfnamefont {M.}~\bibnamefont
  {Pavone}}, \bibinfo {author} {\bibfnamefont {J.~A.}\ \bibnamefont {Keith}}, \
  and\ \bibinfo {author} {\bibfnamefont {E.~A.}\ \bibnamefont {Carter}},\
  }\href {\doibase 10.1021/ar4003174} {\bibfield  {journal} {\bibinfo
  {journal} {Acc. Chem. Res.}\ }\textbf {\bibinfo {volume} {47}},\ \bibinfo
  {pages} {3340} (\bibinfo {year} {2014})}\BibitemShut {NoStop}%
\bibitem [{\citenamefont {Xue}\ \emph {et~al.}(2020)\citenamefont {Xue},
  \citenamefont {Wang},\ and\ \citenamefont {Yang}}]{Xue2020}%
  \BibitemOpen
  \bibfield  {author} {\bibinfo {author} {\bibfnamefont {J.}~\bibnamefont
  {Xue}}, \bibinfo {author} {\bibfnamefont {R.}~\bibnamefont {Wang}}, \ and\
  \bibinfo {author} {\bibfnamefont {Y.}~\bibnamefont {Yang}},\ }\href {\doibase
  10.1038/s41578-020-0221-1} {\bibfield  {journal} {\bibinfo  {journal} {Nat.
  Rev. Mater.}\ }\textbf {\bibinfo {volume} {5}},\ \bibinfo {pages} {809}
  (\bibinfo {year} {2020})}\BibitemShut {NoStop}%
\bibitem [{\citenamefont {DeQuilettes}\ \emph {et~al.}(2015)\citenamefont
  {DeQuilettes}, \citenamefont {Vorpahl}, \citenamefont {Stranks},
  \citenamefont {Nagaoka}, \citenamefont {Eperon}, \citenamefont {Ziffer},
  \citenamefont {Snaith},\ and\ \citenamefont {Ginger}}]{Dane2015}%
  \BibitemOpen
  \bibfield  {author} {\bibinfo {author} {\bibfnamefont {D.~W.}\ \bibnamefont
  {DeQuilettes}}, \bibinfo {author} {\bibfnamefont {S.~M.}\ \bibnamefont
  {Vorpahl}}, \bibinfo {author} {\bibfnamefont {S.~D.}\ \bibnamefont
  {Stranks}}, \bibinfo {author} {\bibfnamefont {H.}~\bibnamefont {Nagaoka}},
  \bibinfo {author} {\bibfnamefont {G.~E.}\ \bibnamefont {Eperon}}, \bibinfo
  {author} {\bibfnamefont {M.~E.}\ \bibnamefont {Ziffer}}, \bibinfo {author}
  {\bibfnamefont {H.~J.}\ \bibnamefont {Snaith}}, \ and\ \bibinfo {author}
  {\bibfnamefont {D.~S.}\ \bibnamefont {Ginger}},\ }\href@noop {} {\bibfield
  {journal} {\bibinfo  {journal} {Science}\ }\textbf {\bibinfo {volume}
  {384}},\ \bibinfo {pages} {683} (\bibinfo {year} {2015})}\BibitemShut
  {NoStop}%
\bibitem [{\citenamefont {DeQuilettes}\ \emph {et~al.}(2016)\citenamefont
  {DeQuilettes}, \citenamefont {Zhang}, \citenamefont {Burlakov}, \citenamefont
  {Graham}, \citenamefont {Leijtens}, \citenamefont {Osherov}, \citenamefont
  {Bulovi{\'{c}}}, \citenamefont {Snaith}, \citenamefont {Ginger},\ and\
  \citenamefont {Stranks}}]{DeQuilettes2016}%
  \BibitemOpen
  \bibfield  {author} {\bibinfo {author} {\bibfnamefont {D.~W.}\ \bibnamefont
  {DeQuilettes}}, \bibinfo {author} {\bibfnamefont {W.}~\bibnamefont {Zhang}},
  \bibinfo {author} {\bibfnamefont {V.~M.}\ \bibnamefont {Burlakov}}, \bibinfo
  {author} {\bibfnamefont {D.~J.}\ \bibnamefont {Graham}}, \bibinfo {author}
  {\bibfnamefont {T.}~\bibnamefont {Leijtens}}, \bibinfo {author}
  {\bibfnamefont {A.}~\bibnamefont {Osherov}}, \bibinfo {author} {\bibfnamefont
  {V.}~\bibnamefont {Bulovi{\'{c}}}}, \bibinfo {author} {\bibfnamefont {H.~J.}\
  \bibnamefont {Snaith}}, \bibinfo {author} {\bibfnamefont {D.~S.}\
  \bibnamefont {Ginger}}, \ and\ \bibinfo {author} {\bibfnamefont {S.~D.}\
  \bibnamefont {Stranks}},\ }\href {\doibase 10.1038/ncomms11683} {\bibfield
  {journal} {\bibinfo  {journal} {Nat. Comm.}\ }\textbf {\bibinfo {volume}
  {7}},\ \bibinfo {pages} {11683} (\bibinfo {year} {2016})}\BibitemShut
  {NoStop}%
\bibitem [{\citenamefont {Liu}\ \emph {et~al.}(2018)\citenamefont {Liu},
  \citenamefont {Zhang}, \citenamefont {Shi}, \citenamefont {Liu},
  \citenamefont {Huang}, \citenamefont {Yun}, \citenamefont {Zeng},
  \citenamefont {Pu}, \citenamefont {Sun}, \citenamefont {Hameiri},
  \citenamefont {Stride}, \citenamefont {Seidel}, \citenamefont {Green},\ and\
  \citenamefont {Hao}}]{Liu2018}%
  \BibitemOpen
  \bibfield  {author} {\bibinfo {author} {\bibfnamefont {X.}~\bibnamefont
  {Liu}}, \bibinfo {author} {\bibfnamefont {Y.}~\bibnamefont {Zhang}}, \bibinfo
  {author} {\bibfnamefont {L.}~\bibnamefont {Shi}}, \bibinfo {author}
  {\bibfnamefont {Z.}~\bibnamefont {Liu}}, \bibinfo {author} {\bibfnamefont
  {J.}~\bibnamefont {Huang}}, \bibinfo {author} {\bibfnamefont {J.~S.}\
  \bibnamefont {Yun}}, \bibinfo {author} {\bibfnamefont {Y.}~\bibnamefont
  {Zeng}}, \bibinfo {author} {\bibfnamefont {A.}~\bibnamefont {Pu}}, \bibinfo
  {author} {\bibfnamefont {K.}~\bibnamefont {Sun}}, \bibinfo {author}
  {\bibfnamefont {Z.}~\bibnamefont {Hameiri}}, \bibinfo {author} {\bibfnamefont
  {J.~A.}\ \bibnamefont {Stride}}, \bibinfo {author} {\bibfnamefont
  {J.}~\bibnamefont {Seidel}}, \bibinfo {author} {\bibfnamefont {M.~A.}\
  \bibnamefont {Green}}, \ and\ \bibinfo {author} {\bibfnamefont
  {X.}~\bibnamefont {Hao}},\ }\href {\doibase
  https://doi.org/10.1002/aenm.201800138} {\bibfield  {journal} {\bibinfo
  {journal} {Adv. Energy Mater.}\ }\textbf {\bibinfo {volume} {8}},\ \bibinfo
  {pages} {1800138} (\bibinfo {year} {2018})}\BibitemShut {NoStop}%
\bibitem [{\citenamefont {Chen}\ \emph
  {et~al.}(2019{\natexlab{b}})\citenamefont {Chen}, \citenamefont {Zhou},
  \citenamefont {Chen}, \citenamefont {Wu}, \citenamefont {Tu}, \citenamefont
  {Liu}, \citenamefont {Huang}, \citenamefont {Ng}, \citenamefont
  {Djurišić},\ and\ \citenamefont {He}}]{Chen2019b}%
  \BibitemOpen
  \bibfield  {author} {\bibinfo {author} {\bibfnamefont {W.}~\bibnamefont
  {Chen}}, \bibinfo {author} {\bibfnamefont {Y.}~\bibnamefont {Zhou}}, \bibinfo
  {author} {\bibfnamefont {G.}~\bibnamefont {Chen}}, \bibinfo {author}
  {\bibfnamefont {Y.}~\bibnamefont {Wu}}, \bibinfo {author} {\bibfnamefont
  {B.}~\bibnamefont {Tu}}, \bibinfo {author} {\bibfnamefont {F.-Z.}\
  \bibnamefont {Liu}}, \bibinfo {author} {\bibfnamefont {L.}~\bibnamefont
  {Huang}}, \bibinfo {author} {\bibfnamefont {A.~M.~C.}\ \bibnamefont {Ng}},
  \bibinfo {author} {\bibfnamefont {A.~B.}\ \bibnamefont {Djurišić}}, \ and\
  \bibinfo {author} {\bibfnamefont {Z.}~\bibnamefont {He}},\ }\href {\doibase
  https://doi.org/10.1002/aenm.201803872} {\bibfield  {journal} {\bibinfo
  {journal} {Adv. Energy Mater.}\ }\textbf {\bibinfo {volume} {9}},\ \bibinfo
  {pages} {1803872} (\bibinfo {year} {2019}{\natexlab{b}})}\BibitemShut
  {NoStop}%
\bibitem [{\citenamefont {Apergi}\ \emph {et~al.}(2020)\citenamefont {Apergi},
  \citenamefont {Brocks},\ and\ \citenamefont {Tao}}]{Apergi2020}%
  \BibitemOpen
  \bibfield  {author} {\bibinfo {author} {\bibfnamefont {S.}~\bibnamefont
  {Apergi}}, \bibinfo {author} {\bibfnamefont {G.}~\bibnamefont {Brocks}}, \
  and\ \bibinfo {author} {\bibfnamefont {S.}~\bibnamefont {Tao}},\ }\href
  {\doibase 10.1103/PhysRevMaterials.4.085403} {\bibfield  {journal} {\bibinfo
  {journal} {Phys. Rev. Materials}\ }\textbf {\bibinfo {volume} {4}},\ \bibinfo
  {pages} {085403} (\bibinfo {year} {2020})}\BibitemShut {NoStop}%
\bibitem [{\citenamefont {Chen}\ \emph {et~al.}(2020)\citenamefont {Chen},
  \citenamefont {Lei}, \citenamefont {Li}, \citenamefont {Yu}, \citenamefont
  {Cai}, \citenamefont {Chiu}, \citenamefont {Rao}, \citenamefont {Gu},
  \citenamefont {Wang}, \citenamefont {Choi}, \citenamefont {Hu}, \citenamefont
  {Wang}, \citenamefont {Li}, \citenamefont {Song}, \citenamefont {Zhang},
  \citenamefont {Qi}, \citenamefont {Lin}, \citenamefont {Zhang}, \citenamefont
  {Islam}, \citenamefont {Maruyama}, \citenamefont {Dayeh}, \citenamefont {Li},
  \citenamefont {Yang}, \citenamefont {Lo},\ and\ \citenamefont
  {Xu}}]{Chen2020}%
  \BibitemOpen
  \bibfield  {author} {\bibinfo {author} {\bibfnamefont {Y.}~\bibnamefont
  {Chen}}, \bibinfo {author} {\bibfnamefont {Y.}~\bibnamefont {Lei}}, \bibinfo
  {author} {\bibfnamefont {Y.}~\bibnamefont {Li}}, \bibinfo {author}
  {\bibfnamefont {Y.}~\bibnamefont {Yu}}, \bibinfo {author} {\bibfnamefont
  {J.}~\bibnamefont {Cai}}, \bibinfo {author} {\bibfnamefont {M.~H.}\
  \bibnamefont {Chiu}}, \bibinfo {author} {\bibfnamefont {R.}~\bibnamefont
  {Rao}}, \bibinfo {author} {\bibfnamefont {Y.}~\bibnamefont {Gu}}, \bibinfo
  {author} {\bibfnamefont {C.}~\bibnamefont {Wang}}, \bibinfo {author}
  {\bibfnamefont {W.}~\bibnamefont {Choi}}, \bibinfo {author} {\bibfnamefont
  {H.}~\bibnamefont {Hu}}, \bibinfo {author} {\bibfnamefont {C.}~\bibnamefont
  {Wang}}, \bibinfo {author} {\bibfnamefont {Y.}~\bibnamefont {Li}}, \bibinfo
  {author} {\bibfnamefont {J.}~\bibnamefont {Song}}, \bibinfo {author}
  {\bibfnamefont {J.}~\bibnamefont {Zhang}}, \bibinfo {author} {\bibfnamefont
  {B.}~\bibnamefont {Qi}}, \bibinfo {author} {\bibfnamefont {M.}~\bibnamefont
  {Lin}}, \bibinfo {author} {\bibfnamefont {Z.}~\bibnamefont {Zhang}}, \bibinfo
  {author} {\bibfnamefont {A.~E.}\ \bibnamefont {Islam}}, \bibinfo {author}
  {\bibfnamefont {B.}~\bibnamefont {Maruyama}}, \bibinfo {author}
  {\bibfnamefont {S.}~\bibnamefont {Dayeh}}, \bibinfo {author} {\bibfnamefont
  {L.~J.}\ \bibnamefont {Li}}, \bibinfo {author} {\bibfnamefont
  {K.}~\bibnamefont {Yang}}, \bibinfo {author} {\bibfnamefont {Y.~H.}\
  \bibnamefont {Lo}}, \ and\ \bibinfo {author} {\bibfnamefont {S.}~\bibnamefont
  {Xu}},\ }\href@noop {} {\bibfield  {journal} {\bibinfo  {journal} {Nature}\
  }\textbf {\bibinfo {volume} {577}},\ \bibinfo {pages} {209} (\bibinfo {year}
  {2020})}\BibitemShut {NoStop}%
\bibitem [{\citenamefont {Jinnouchi}\ \emph {et~al.}(2019)\citenamefont
  {Jinnouchi}, \citenamefont {Lahnsteiner}, \citenamefont {Karsai},
  \citenamefont {Kresse},\ and\ \citenamefont {Bokdam}}]{Jinnouchi2019}%
  \BibitemOpen
  \bibfield  {author} {\bibinfo {author} {\bibfnamefont {R.}~\bibnamefont
  {Jinnouchi}}, \bibinfo {author} {\bibfnamefont {J.}~\bibnamefont
  {Lahnsteiner}}, \bibinfo {author} {\bibfnamefont {F.}~\bibnamefont {Karsai}},
  \bibinfo {author} {\bibfnamefont {G.}~\bibnamefont {Kresse}}, \ and\ \bibinfo
  {author} {\bibfnamefont {M.}~\bibnamefont {Bokdam}},\ }\href@noop {}
  {\bibfield  {journal} {\bibinfo  {journal} {Phys. Rev. Lett.}\ }\textbf
  {\bibinfo {volume} {122}},\ \bibinfo {pages} {225701} (\bibinfo {year}
  {2019})}\BibitemShut {NoStop}%
\end{thebibliography}
\end{document}